\documentclass[%
 reprint,
superscriptaddress,
 amsmath,amssymb,
 aps,
]{revtex4-1}

\usepackage{graphicx}
\usepackage{dcolumn}
\usepackage{bm}
\usepackage{xcolor,colortbl}
\usepackage{float}
\usepackage[colorlinks=true, allcolors=blue]{hyperref}
\usepackage[colorinlistoftodos,textsize=tiny]{todonotes}

\usepackage{booktabs} 
\definecolor{lightgray}{RGB}{211,211,211}
\definecolor{Red}{RGB}{153,0,0}
\definecolor{Magenta}{RGB}{153,0,153}
\definecolor{Orange}{RGB}{204,102,0}
\definecolor{Teal}{RGB}{0,153,153}
\definecolor{LightBlue}{RGB}{0,128,255}
\definecolor{LightGreen}{RGB}{0,153,76}

\usepackage{mathtools}

\begin{document}

\preprint{APS/123-QED}

\title{Stochastic force dynamics of the model microswimmer \emph{Chlamydomonas reinhardtii}: Active forces and energetics}

\author{Corbyn Jones}
\thanks{equally contributing authors}
\affiliation{Department of Physics,  California State University Fullerton, CA 92831 USA}

\author{Mauricio Gomez}
\thanks{equally contributing authors}
\affiliation{Department of Physics,  California State University Fullerton, CA 92831 USA}

\author{Ryan M. Muoio}
\thanks{equally contributing authors}
\affiliation{Department of Physics,  California State University Fullerton, CA 92831 USA}

\author{Alex Vidal}
\affiliation{Department of Computer Science, California State University Fullerton, CA 92831 USA}

\author{Anthony Mcknight}
\affiliation{Department of Biomedical Engineering, University of Virginia, Charlottesville, VA 22904 USA. }

\author{Nicholas D. Brubaker}
\affiliation{Department of Mathematics, California State University Fullerton, CA 92831 USA}

\author{Wylie W. Ahmed}
\email[correspondence: ]{wahmed@fullerton.edu }
\affiliation{Department of Physics,  California State University Fullerton, CA 92831 USA}


\date{\today}

\begin{abstract}
We study the stochastic force dynamics of a model microswimmer (\emph{Chlamydomonas reinhardtii}), using a combined experimental, theoretical, and numerical approach.  While swimming dynamics have been extensively studied using hydrodynamic approaches, which infer forces from the viscous flow field, we directly measure the stochastic forces generated by the microswimmer using an optical trap via the photon momentum method.  We analyze the force dynamics by modeling the microswimmer as a self-propelled particle, \emph{\`{a} la} active matter, and analyze it's energetics using methods from stochastic thermodynamics.  We find complex oscillatory force dynamics and power dissipation on the order of $10^6$ $k_B T / s$ ($\sim$ fW).
\end{abstract}

\maketitle


\section{Introduction}

Swimming at the microscopic scale has long attracted the interest of biologists, physicists, and applied mathematicians~\cite{lauga2012microswimmers, purcell1977life}.  The self-propulsion of microorganisms through viscous fluids at low Reynolds number is an essential aspect of life~\cite{lauga2009hydrodynamics,cisneros2010fluid,elgeti2015physics,guasto2012fluid,koch2011collective}.  At the most basic level, this involves a swimmer physically interacting with its environment to create directed motion.  A widely studied model microswimmer is the \emph{Chlamydomonas reinhardtii}, which is a unicellular biflagellate alga that uses a breaststroke motion to pull itself through its environment~\cite{jeanneret2016brief,goldstein2015green}.  \emph{Chlamydomonas} have been studied extensively by biologists~\cite{harris2001chlamydomonas,harris2009chlamydomonas, merchant2007chlamydomonas} long before physicists became interested~\cite{pedley1992hydrodynamic}.

To understand the swimming dynamics of the \emph{Chlamydomonas}, hydrodynamic approaches have been used extensively to understand the flow fields around the swimmer and infer the forces generated by them~\cite{elgeti2015physics}.  A collection of impactful experimental and theoretical studies have uncovered complex swimming dynamics~\cite{polin2009chlamydomonas, drescher2010direct}, flagellar waveforms~\cite{bayly2011propulsive,kurtuldu2013flagellar}, enhanced diffusion~\cite{leptos2009dynamics, brun2019effective}, synchronization~\cite{brumley2014flagellar,niedermayer2008synchronization,uchida2011generic, wan2014lag, guo2020intracellular}, and fluctuations~\cite{wan2014lag, wan2018time}.  Studies typically analyze the swimmer motion and the surrounding fluid velocity field to infer forces using Stokeslet models~\cite{lighthill1952squirming,hancock1953self, blake1971note, lauga2009hydrodynamics}.  However, direct investigation of the force dynamics has been more elusive.  

Here, we present a study of direct measurement of the stochastic forces generated by a \emph{Chlamydomonas} and interpret the dynamics via modeling and simulations of a self-propelled particle. Optical trapping to manipulate and study microswimmers is not new~\cite{ashkin1987optical,ashkin1987optical2, block1989compliance,min2009high}, however, \emph{in-situ }force calibration is typically challenging~\cite{fischer2007calibration,hendricks2012force,blehm2013vivo, ahmed2018active}. Here, direct force measurement is possible due to recent advances in optical trap calibration, known as the photon momentum method~\cite{farre2010force,farre2017beyond}, that measures force via changes in photon momentum of the trapping beam and does not require \emph{a priori} knowledge of the object shape, refractive index, a synthetic handle, or external calibration as used traditionally for quantitative force measurement~\cite{oddershede2012force, chattopadhyay2006swimming, stellamanns2014optical, mccord2005analysis, fallman2004optical}. This approach allows study of the stochasticity of forces given the high resolution in force and sampling rate (sub-pN and 50 kHz, respectively). This approach is particularly interesting for studies of stochastic thermodynamics where fluctuations are important because it allows direct access to force without assuming an underlying model (e.g.~a linear spring), does not require averaging, and is not limited by the precision of image acquisition or correlation techniques~\cite{seifert2012stochastic, farre2017beyond,ouellette2006quantitative}.  Since the photon momentum method does not rely on trap linearity it allows large force fluctuations in the nonlinear range to be captured, unlike traditional techniques for calibration such as equipartition, Stokes drag, or active-passive approaches~\cite{pesce2020optical}.   By treating the swimmer as an ``active particle'' we quantify its non-equilibrium activity using tools from stochastic thermodynamics~\cite{seifert2012stochastic, harada2005equality,shinkai2014energetics, eldeen2020quantifying}.  We find that \emph{Chlamydomonas} exhibit complex oscillatory force dynamics with magnitude of tens of pNs, and rotational dynamics of 1-2 Hz, that can be characterized using a self-propelled particle model.  Measurements and theory suggest the power dissipated by single swimmers to be on the order of $10^6$ $k_B T/s$ ($\sim$ fW).

\section{Theoretical Model and Experimental Methods}

\subsection{Equation of motion}
We model the stochastic motion and active force dynamics of the optically trapped \emph{Chlamydomonas} (in the two-dimensional horizontal plane) with the overdamped Langevin equation~\cite{fodor2018statistical,sekimoto1998langevin}. That is, the position $\mathbf{r}(t) \in \mathbb{R}^2$ of the \emph{Chlamydomonas} (or active particle) is governed by
\begin{equation} 
    \gamma\dot{\mathbf{r}}+\kappa \mathbf{r}=\gamma \mathbf{u}+\sqrt{2D}\gamma \boldsymbol\eta,
    \label{eqn:EoM}
\end{equation}
which balances the deterministic frictional and optical trap forces with the random active and thermal forces. In \eqref{eqn:EoM}, $\kappa$ is the optical trap stiffness, $\gamma$ is the friction coefficient of the Stokes' drag, and $D$ is the thermal diffusion coefficient of the zero-mean, $\delta$-correlated Gaussian white noise process $\boldsymbol\eta$. Also, $\mathbf{u}$ is the intrinsic  self-propulsion velocity of the active particle that can take on many forms depending on the underlying model of the active particle such as Active Brownian Particle (ABP) or Active Ornstein-Uhlenbeck Particle (AOUP)~\cite{fodor2018statistical,romanczuk2012active}.  In our analytic approach we use the AOUP model as discussed in the appendix. From this equation of motion we derive analytic quantities such as the force spectrum (see appendix) and perform numerical simulations of particle trajectories.  

\subsection{Numerical Simulations}
To simulate individual trajectories we discretize~\eqref{eqn:EoM} using the Euler-Maruyama method~\cite{volpe2014simulation,higham2001algorithmic} over uniform time steps of size $\Delta t$, which produces the iterative procedure
\begin{equation}
    \mathbf{r}_i = \mathbf{r}_{i-1} - \frac{\kappa}{\gamma}\mathbf{r}_{i-1}\Delta t + \mathbf{u}_{i-1}\Delta t + \sqrt{2D\Delta t}\,\mathbf{Z}_{i-1},
    \label{eqn:DiscreteTrajectory}
\end{equation}
for $i = 1,2,\ldots.$ The subscripts indicate the time step of the corresponding quantities, and $\mathbf{Z}_i$ is a normally distributed multivariate random variable that has zero mean and covariance equal to the identity matrix; see appendix for the derivation of equation \ref{eqn:DiscreteTrajectory}.

The forces due to friction, the trapping potential, active processes, and thermal fluctuations at each time step are then defined by 
\begin{align*}
    \mathbf{F}_{\mathrm{fric},i} &= -\gamma\frac{\Delta\mathbf{r}_i}{\Delta t},\\
    \mathbf{F}_{\mathrm{trap},i} &= -\kappa\mathbf{r}_i,\\
    \mathbf{F}_{\mathrm{act},i} &= \gamma\mathbf{u}_i,\\
    \mathbf{F}_{\mathrm{th},i} &= \sqrt{\frac{2D}{\Delta t}}\mathbf{Z}_{i},
\end{align*}
and induce the respective work increments
\begin{align*}
    \Delta W_{\mathrm{fric},i} &= \mathbf{F}_{\mathrm{fric},i}\circ\Delta\mathbf{r}_i,\\
    \Delta W_{\mathrm{trap},i} &= \mathbf{F}_{\mathrm{trap},i}\circ\Delta\mathbf{r}_i,\\
    \Delta W_{\mathrm{act},i} &= \mathbf{F}_{\mathrm{act},i}\circ\Delta\mathbf{r}_i,\\
    \Delta W_{\mathrm{th},i} &= \mathbf{F}_{\mathrm{th},i}\circ\Delta\mathbf{r}_i,
\end{align*}
where the symbol $\circ$ denotes the dot product with Stratonovich convention~\cite{gardiner1985handbook}, e.g.~$\mathbf{F}(t^\prime) = [ \mathbf{F}(t_i) + \mathbf{F}(t_{i-1}) ] / 2$. In this discrete setting the forces and their induced work increments are mathematically well-defined; however, as the time step $\Delta t \to 0^+$, the frictional and thermal quantities diverge. Experimentally, only the optical trap and frictional terms are directly accessible, which limits our ability to make direct quantitative comparisons for continuous time. We circumvent this issue by making calculations and comparisons with discrete quantities for a time step $\Delta t$ defined by the precision of the experiments.

\begin{figure*}[ht]
\includegraphics[width=1\textwidth]{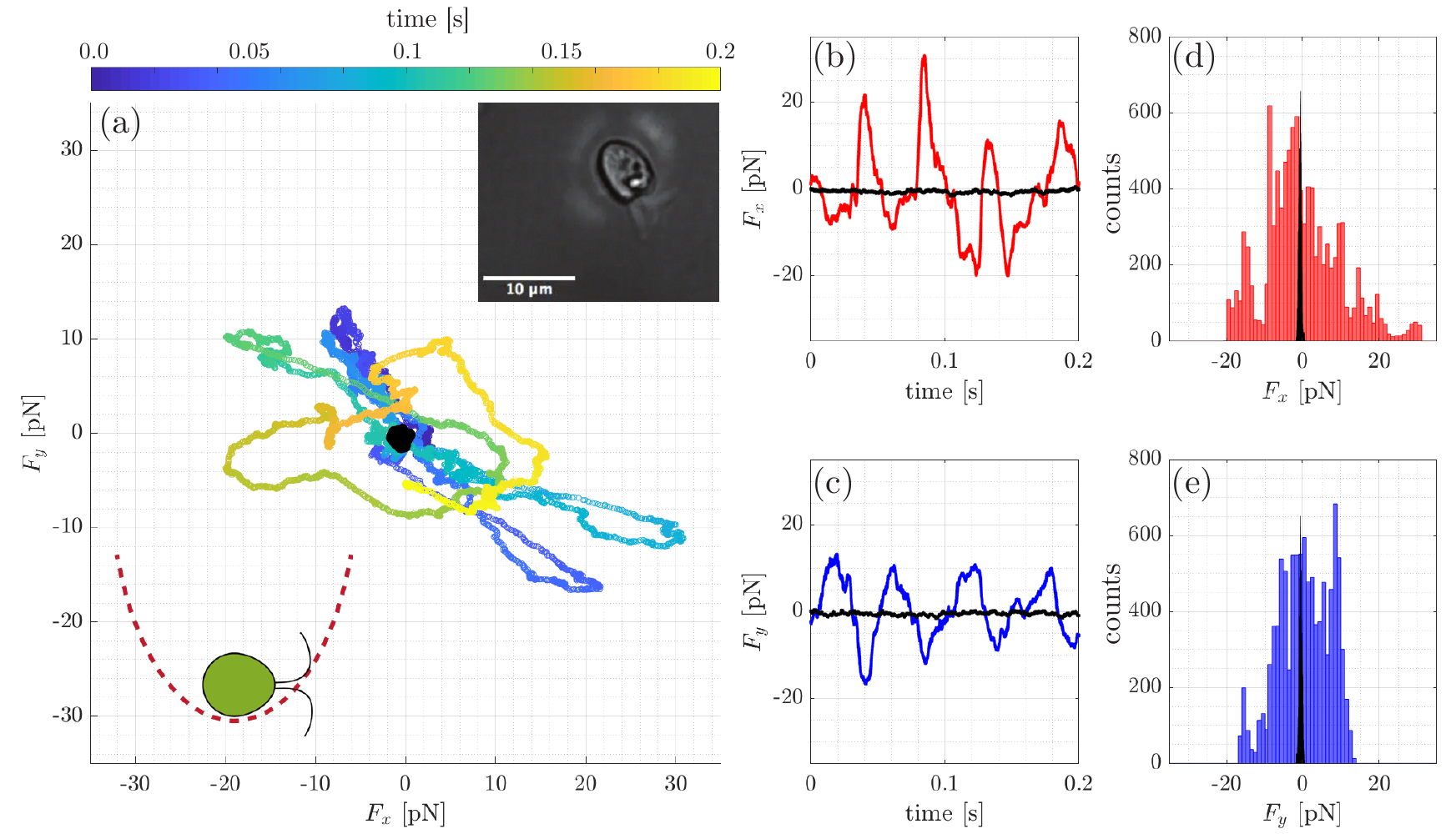}
\caption{\label{fig:char}Representative force dynamics of a single swimming \emph{Chlamydomonas}. (a) Force trajectory plot.  Upper right inset shows representative image of trapped swimmer.  Lower left inset depicts cartoon of swimmer in harmonic trapping potential. (b,c) Force components as a function of time. d,e) Histograms of forces measured in b,c. The black dataset in each panel shows the force dynamics of a passive particle for comparison to thermal fluctuations.}
\end{figure*}

\subsection{Sample Preparation and Optical Tweezer Measurements}
\emph{Chlamydomonas reinhardtii} were purchased from Carolina Scientific (item \#152030) and used within 48 hrs of arrival.  A 20 $\mu$L droplet of stock solution containing \emph{Chlamydomonas} was sandwiched in a sample chamber made from a glass slide and a coverslip (Fisher Scientific) with Dow Corning vacuum grease used as a spacer.  A Nikon TE2000 with a 60x/1.2NA water-immersion objective and Hamamatsu ORCA-Flash4.0 V2 was used for microscopy.  The optical tweezer system (Impetux Optics S.L.) includes the optical trap, piezo stage positioning, and force detection.  The 60x objective focuses the near-infrared fiber laser (1064 nm, IPG-YLR-10, IPG Photonics) to create the optical trap. Force detection and laser tracking interferometry is done using the photon momentum method (PMM)~\cite{farre2010force, gieseler2020optical} implemented with a 1.4NA oil immersion condenser and a position sensitive sensor that is digitized at 50 kHz. For the PMM method force calibration to be accurate it is critical to use a condensing objective with higher numerical aperture than the trapping objective and to minimize scatter of light through the sample~\cite{farre2010force,jun2014calibration}.  All control of experimental hardware and data acquisition was done using Labview (National Instruments).

\subsection{Data Analysis}
All data analysis of experiments and simulations was completed in MATLAB.  To calculate the force spectrum, the power spectrum of a finite force signal, $\mathbf{F}(t)$ sampled at 50 kHz, was estimated using Welch's method~\cite{welch1967use} with a Hamming window.  This approach was used for both experimental data and numerically simulated data.  Fitting of the force spectrum to our analytic model was done using nonlinear least squares~\cite{coleman1996interior}.  For calculations of work fluctuations in the time domain, data was downsampled by a factor of 10 to reduce high-frequency noise, for a resulting sampling frequency of 5 kHz.

\begin{figure*}[t]
\includegraphics[width=1\textwidth]{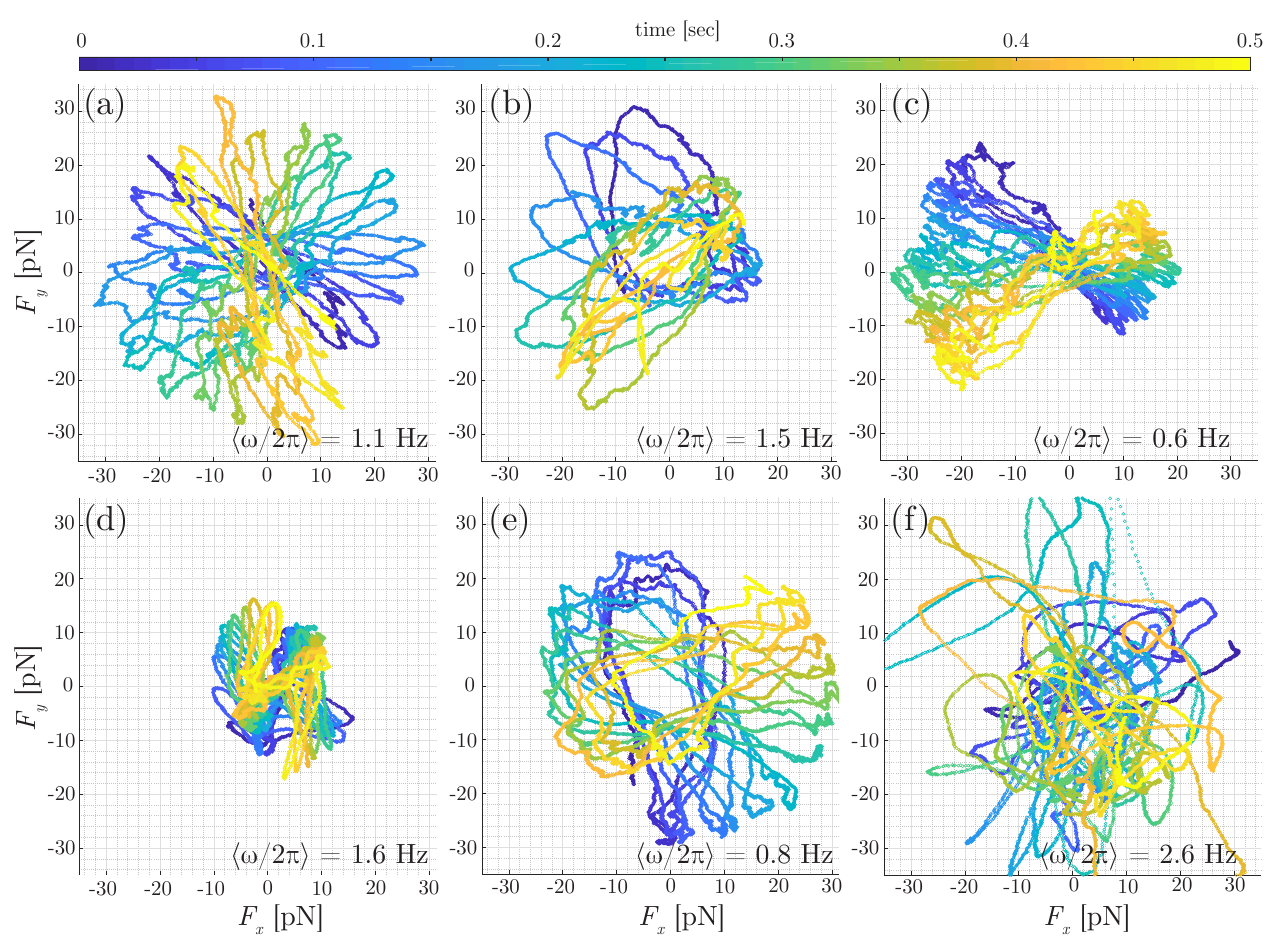}
\caption{\label{fig:gallery}Gallery of force dynamics of an optically trapped \emph{Chlamydomonas}.  A wide array of patterns in the force trajectories are observed, and six representative examples are shown here (a-f).  The overall rotational motion, $\langle \omega / 2 \pi \rangle$, is compatible with previous studies~\cite{goldstein2015green}. }
\end{figure*}

\section{Results and Discussion}

\subsection{Characterizing force dynamics}

Most studies of \emph{Chlamydomonas reinhardtii} swimming measure the position/velocity of the swimmer itself or the fluid field surrounding it~\cite{racey1981quasi,drescher2010direct,bayly2011propulsive, guasto2012fluid}.  These studies provide a wealth of information, particularly on swimming dynamics~\cite{polin2009chlamydomonas, guasto2010oscillatory, cortese2020control}, flagellar waveforms~\cite{kurtuldu2013flagellar, wan2014rhythmicity}, and synchronization~\cite{goldstein2011emergence,brumley2014flagellar, quaranta2015hydrodynamics}. We take an alternative approach where we directly measure the force generated by the swimmer as applied to the optical trap.  This is possible due to a recently developed force calibration technique, called the photon momentum method (PMM)~\cite{farre2010force,farre2017beyond}, that does not require \emph{a priori} knowledge of the trapped particle or the surrounding fluid bath.  One might expect that inferred forces from fluid mechanics approaches~\cite{guasto2012fluid,elgeti2015physics} and direct measurement of the force generated by a swimmer would be compatible, since they should be related by Newton's 3rd law.  Our direct measurements of force confirm  this hypothesis as we describe in this section.

The characteristic force dynamics of a \emph{Chlamydomonas} in an optical trap is shown in Fig.~\ref{fig:char}, where only a short snippet (0.2 s) is shown for visibility purposes.  By direct force measurement it is evident the swimmer exhibits oscillatory force dynamics in both $x$ and $y$ directions due to its breast stroke motion (Fig.~\ref{fig:char}a-c).  Looking at a short portion of the trajectory, the first interesting feature is the forward/backward asymmetry of the forces, most evident in Fig.~\ref{fig:char}b in the first 0.1s, where the swimmer is dominantly aligned in the $x$-direction and the forward (positive) beating force is larger than the backward (negative) force.  This is supported by the corresponding force histogram (Fig.~\ref{fig:char}d) where there is a positive peak of $\sim$30 pN and a negative peak of $\sim$-15 pN.  For comparison to fluids approaches, if one assumes simple Stokes friction ($F = \gamma v$), then the force measured predicts forward/backward swimming velocities of $\sim$400/200 $\mu$m/s, which is in close agreement with reported values~\cite{kurtuldu2013flagellar,racey1981quasi,guasto2010oscillatory}.  

The second interesting feature to note is the observed force trajectory is not simply a forward/backward motion through the center of the trap, but rather a complex oscillation that drifts around the origin over time.  Note that since the optical trap applies no torque, the swimmer is free to rotate.  We observe a wide variety of rotational patterns as shown in a gallery of force trajectories in Fig.~\ref{fig:gallery}.  Similar dynamics are observed for all swimmers, meaning a general rotation is always present, but we note the finer features of the force trajectory show great complexity.  Focusing on the average rotation of the forward/backward swimming axis, a spectrogram analysis reveals an angular velocity, $\langle \omega / 2\pi \rangle$, as labeled in Fig.~\ref{fig:gallery}. This angular velocity extracted from the force trajectories is consistent with the \emph{Chlamydomonas'} well-documented spinning motion of $\sim1-2$ Hz from motion-tracking studies~\cite{ruffer1985high,harris2009chlamydomonas, goldstein2015green, cortese2020control}.

\begin{figure}[t]
\includegraphics[width=0.5\textwidth]{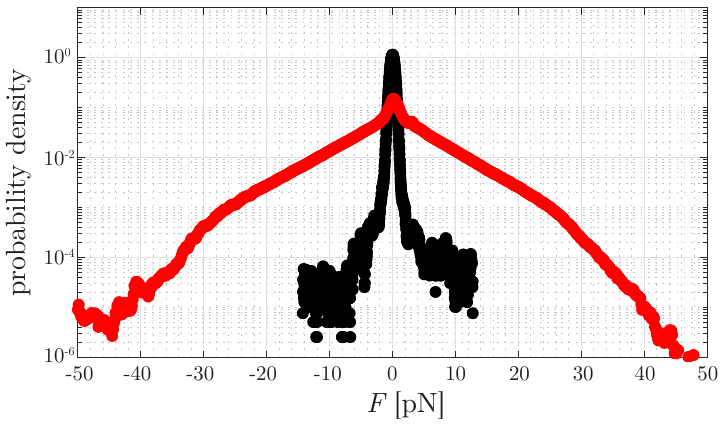}
\caption{\label{fig:forceHist} Probability distribution of forces measured by the optical trap for a \emph{Chlamydomonas} (red) and a thermally fluctuating particle (black).  This distribution is calculated from $n=24$ experiments.}
\end{figure}

Overall, the force dynamics in terms of peak values, oscillations, and rotation are in agreement with swimming dynamics from previous fluids studies~\cite{goldstein2015green}.  This shows that the optical trap (with PMM calibration) is a viable technique to study the force dynamics of a microswimmer. Further, direct force measurement provides several advantages --- e.g. model independent force information (at sub-pN resolution) and high-temporal resolution (50 kHz) which make this approach ideal for studying the stochasticity of microswimmer generated forces.  For instance, the probability distribution of forces reveals much about the underlying processes occurring. A thermally fluctuating particle exhibits Gaussian force fluctuations characteristic of thermal noise as shown in black in Fig.~\ref{fig:forceHist}.  The force fluctuations of the \emph{Chlamydomonas} are non-Gaussian and have a much wider variance due to the active non-thermal forces generated by the beating dynamics of the swimmer (Fig.~\ref{fig:forceHist}, red).  The stark contrast between a passive Brownian particle and a microscopic swimmer shown in Fig.~\ref{fig:forceHist} is due to consumption/dissipation of non-thermal energy by the swimmer, which drives it far from equilibrium.

\color{black}

In the remainder of this paper, we study the stochastic force dynamics of \emph{Chlamydomonas} in an effort to disentangle the active forces generated by the swimmer itself and the thermal forces coming from the fluid bath. We treat the swimmer as an ``active particle'' that uses internal sources of energy to generate self-propulsion~\cite{bechinger2016active}.  In addition, the swimmer experiences forces from thermal fluctuations, friction, and the harmonic trapping potential as outlined in the modeling section above.  Our goal is to quantitatively characterize the activity of the swimmer and explore its energetics.

\subsection{The force spectrum and characterizing activity}

\color{black}
\begin{figure}[t!]
\includegraphics[width=0.5\textwidth]{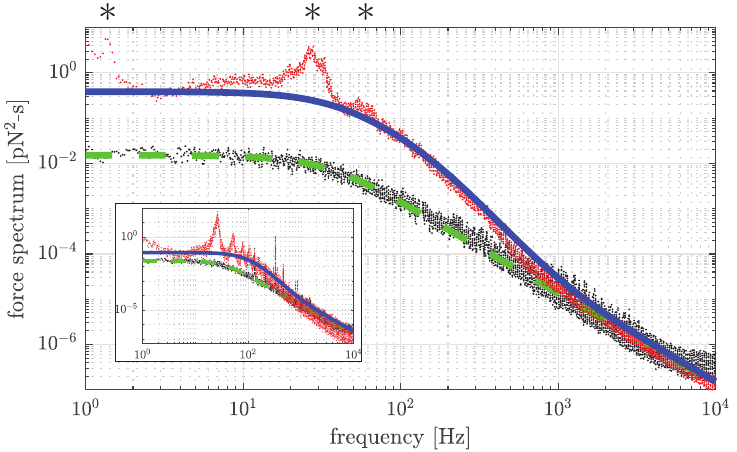}
\caption{\label{fig:PSD} The average total force spectrum of an actively swimming \emph{Chlamydomonas} (red) and the thermal force spectrum (black).  Asterisks ($\ast$) indicate local peaks in the force spectrum at approximately 1.5, 25, and 50 Hz.  Blue solid line and green dashed line are theoretical fits to the analytic model (equation \ref{eqn:psd}).  Inset shows an example of a single measurement that also exhibits local peaks. }
\end{figure}
\color{black}

We use the force spectrum, a recently developed approach~\cite{gallet2009power,guo2014probing,ahmed2018active,bohec2019distribution,eldeen2020quantifying}, to quantify the non-equilibrium force dynamics of the swimmer as an active particle.  The force spectrum is the power spectral density of the stochastic forces measured by the optical trap.  In experiments, we access the force spectra by estimating the power spectral density using Fourier transform methods~\cite{welch1967use}.  In the theoretical model, we calculate the force spectra (equation~\ref{eqn:psd}, derivation in appendix) analytically from the equation of motion,
\color{black}
\begin{equation}
\begin{split}
S\!_{f\!f}(\omega) = & \left(2\kappa^{2}D-\frac{2\tau\kappa^{2}v_{0}^{2}}{\mu^{2}\tau^{2}-1}\right)
\frac{1}{\mu^{2}+\omega^{2}}
\\ & +\frac{2\tau\kappa^{2}v_{0}^{2}}{(\mu^{2}\tau^{2}-1)}
\frac{1}{\tau^{-2}+\omega^{2}},
\end{split}
\label{eqn:psd}
\end{equation}
where $\omega$ is frequency in rad/s, $\mu = \kappa/\gamma$, $\tau$ is the persistence time of the active force, $D$ is the thermal diffusion coefficient, and $v_{0}$ is the characteristic strength of the active velocity.  The majority of parameters in equation~\ref{eqn:psd} are determined by our physical system and the trapped object including: the trap stiffness ($\kappa$), the friction coefficient ($\gamma$), and the thermal diffusion coefficient ($D$).  Thus, equation~\ref{eqn:psd} has only two free parameters that describe the active process: $v_0$, which quantifies the amplitude, and $\tau$, which quantifies the timescale.  In relation to the equation \ref{eqn:EoM}, $v_0 $ is the the average active speed from the AOUP model.

In Fig.~\ref{fig:PSD} we plot the average force spectra of a swimmer (red) and a passive particle (black).  The passive particle is the same size and shape as the swimmer, and thus provides an equilibrium fluctuation baseline (as verified using a dead swimmer).  In both cases, a typical Lorentzian-like shape is evident with a low frequency plateau and a high frequency scaling of $f^{-2}$ expected for thermal fluctuations.  This suggests that for $f > 10^3$ Hz that fluctuations are dominantly thermal.  However, below $10^3$ Hz the two force spectra diverge, clearly showing that the swimmer (red) exhibits force fluctuations greater than a passive particle (black).  The separation between these two force spectra are due to the active forces from the swimmer.  Additionally, the average force spectra of the swimmer exhibits several local peaks, which when averaged over 24 \emph{Chlamydomonas} occur at roughly 1.5, 25, and 50 Hz marked by ($\ast$) in Fig.~\ref{fig:PSD}. These local peaks are not captured by our simple analytic model but are explored later via simulations.

\begin{figure}[t]
\includegraphics[width=0.5\textwidth]{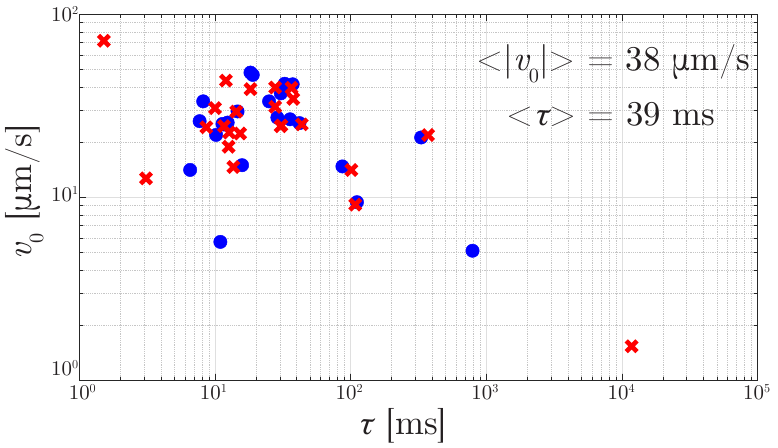}
\caption{\label{fig:Param} Fitting parameters from the analytic model.  $v_0$ represents the amplitude of the active process and $\tau$ the persistence time. Symbols indicate $x$-direction ($\boldsymbol{\times}$) and $y$-direction ($\bullet$).}
\end{figure}

To characterize the measured force fluctuations, we fit equation~\ref{eqn:psd} to each individual force spectra to obtain the two parameters that quantify activity, $v_0$ and $\tau$, which characterize the amplitude of activity and its timescale, respectively.  The extracted parameters for $x$ and $y$ direction are shown in Fig.~\ref{fig:Param} for all experiments.  On average, we find the active speed is $\langle \vert v_0 \vert \rangle = 38$ $\mu$m/s and the active timescale is $\langle \tau \rangle = 39$ ms. These extracted parameters are within the expected range for freely swimming \emph{Chlamydomonas}~\cite{leptos2009dynamics} but biased towards the lower end, likely because our swimmer is trapped.  Nonetheless the agreement is quite striking, given the two experimental methods have little in common, where the fluids approach tracks the position of a free swimmer and our approach measures the force generated by an optically trapped swimmer and fits an analytic model.  It is interesting to note that the extracted $\tau$ does not correspond with the timescale of passive rotational diffusion as it does for active colloids that are well-described by AOUP or ABP models~\cite{fodor2018statistical}.  This suggests $\tau$ is a unique property of the swimmer and is likely related to the flagellar beating and its asymmetry~\cite{polin2009chlamydomonas, cortese2020control}, which would cause decorrelation of the active forces due to rotation. Interestingly, tuning the rotational diffusion independently of thermal fluctuations could allow optimization of swimming trajectories~\cite{fernandez2020feedback}.

\begin{figure}[t]
\includegraphics[width=.5\textwidth]{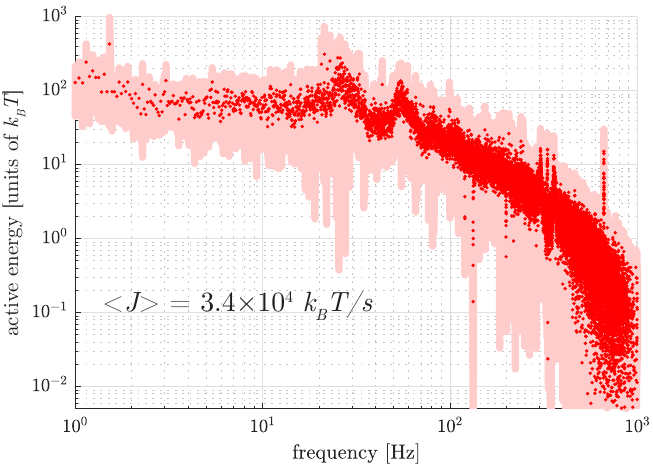}
\caption{\label{fig:Non-eq} The average active energy spectrum quantifies the non-thermal energetic fluctuations of the microswimmer.  Integrating this spectrum provides an estimate of the energy dissipation rate, $\langle J \rangle$, via the Harada-Sasa equality~\cite{harada2005equality}. (shaded region indicates S.E.M.)}
\end{figure}

To isolate the non-equilibrium activity we calculate the average active energy spectrum, $E_{\mathrm{act}} = \langle \vert \tilde{\mathbf{F}}_{\mathrm{tot}} \vert ^2 \rangle  /  \langle \vert \tilde{\mathbf{F}}_{\mathrm{th}} \vert^2 \rangle - 1$, where $\langle \vert \tilde{\mathbf{F}}_{\mathrm{tot}} \vert ^2 \rangle$ is the total force spectrum of the swimmer and $\langle \vert \tilde{\mathbf{F}}_{\mathrm{th}} \vert^2 \rangle$ is the spectrum of thermal forces, and $\tilde{}$ indicates the frequency-domain~\cite{fodor2016nonequilibrium,eldeen2020quantifying}.  The active energy spectrum quantifies the energetic fluctuations due to non-thermal processes and thus characterizes the energy injected into the system by the swimmer (Fig.~\ref{fig:Non-eq}).  Integrating the active energy spectrum provides an estimate for the energy dissipation rate, $\langle J \rangle$, due to active processes~\cite{harada2005equality, eldeen2020quantifying}.  Averaged over all swimmers and all time, we find $\langle J \rangle = 3.4 \times 10^4$ $k_B T$/s ($\sim 0.1$ fW).  Our average energy dissipation rate is significantly lower than the average power dissipated during a \emph{Chlamydomonas} beat cycle ($\sim 4$ fW) as measured from viscous dissipation~\cite{guasto2010oscillatory}.  This could be due to our very different approaches --- measuring force fluctuations vs.~fluid velocity field --- or could be due to other experimental differences.  We offer two factors that may contribute to our measured lower average dissipation: (1) we average over all microswimmers, which includes lower activity samples that may bias the average towards lower dissipation; (2) we average over all time, which includes periods of high, low, and no swimming activity.  To further investigate (1) and (2) we analyze the force dynamics of individual \emph{Chlamydomonas} in the time domain in the following section.

\subsection{Trajectory-level fluctuations}
\begin{figure*}
\includegraphics[width=1\textwidth]{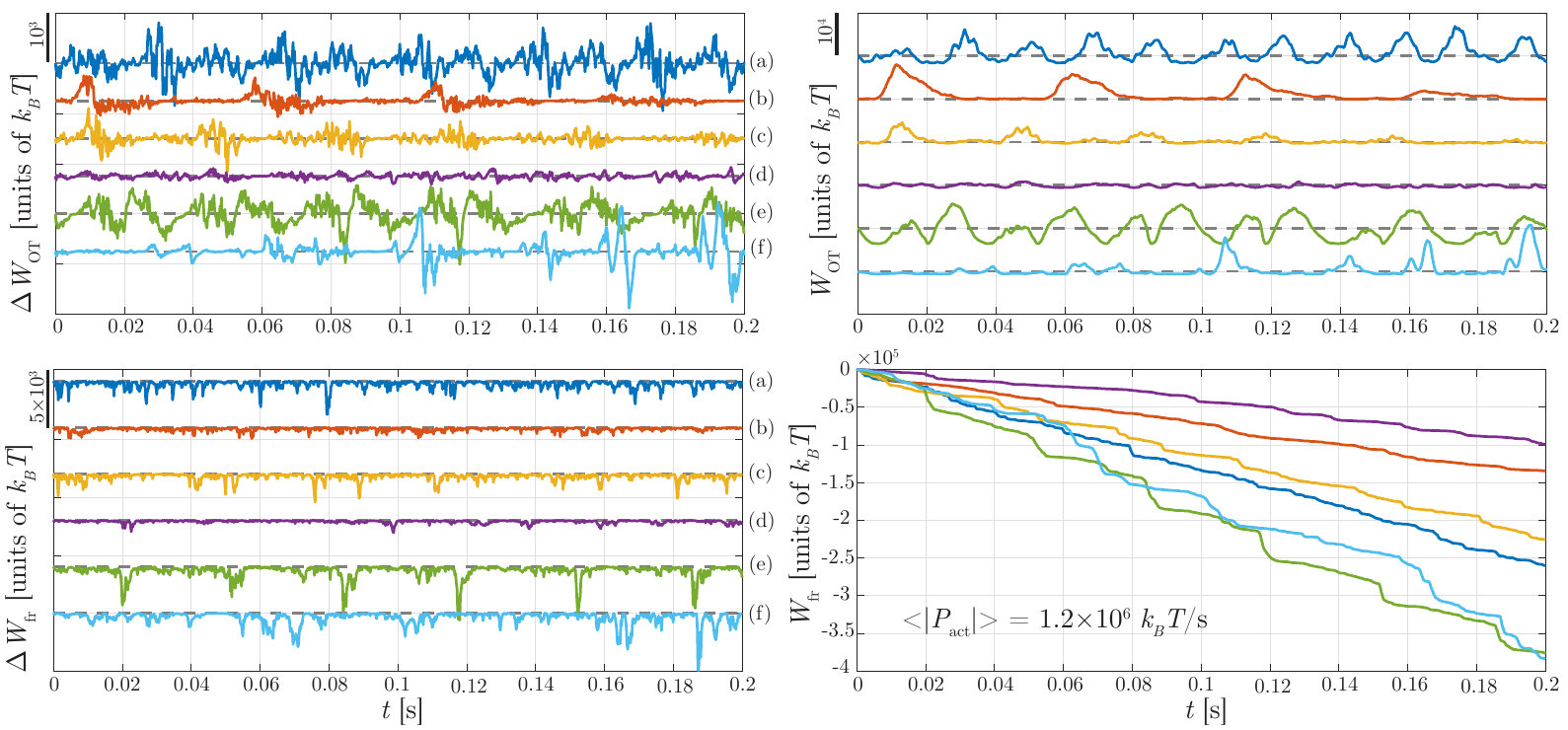}
\caption{\label{fig:work} Work fluctuations (left column) and accumulated work (right column) for trajectories shown in the force gallery (Fig.~\ref{fig:gallery}).  Individual microswimmers (a-f) exhibit significant variation in their dynamics as shown by their time-series dynamics.  This is particularly evident in $W_{\mathrm{fr}}$ where the power dissipated in the observed interval varies by a factor of 4 from $\sim2$ fW (d, purple) to $\sim8$ fW (e, green).}
\end{figure*}

Direct force measurement with an optical trap provides high-resolution trajectories to analyze fluctuations.  We characterize the energetic fluctuations of the swimmers shown in the force gallery (Fig.~\ref{fig:gallery}) by calculating work fluctuations.  Two types of work fluctuations are accessible in our experiments:~(1)~work done by the optical trap; and~(2)~work done by friction.  To access these we assume the optical trap has a linear force-displacement relationship and that the swimmer experiences low-Reynolds Stokes friction.  This assumption allows us to calculate the incremental work done by the optical trap as $\Delta W_{\mathrm{OT}} = \mathbf{F}_{\mathrm{OT}} \circ \Delta \mathbf{r}$, where $\mathbf{F}_{\mathrm{OT}}$ is the force measured by the optical trap, $\Delta \mathbf{r}$ is the incremental displacement, and $\circ$ indicates Stratonovich convention.  Similarly, we can calculate the incremental work done by friction as $\Delta W_{\mathrm{fr}} = \mathbf{F}_{\mathrm{fr}} \circ \Delta \mathbf{r}$, where $\mathbf{F}_{\mathrm{fr}} = - \gamma \dot{\mathbf{r}}$ is the friction force. 

The incremental work fluctuations are plotted in Fig.~\ref{fig:work}(left) where the labels (a)-(f) correspond to the trajectories shown in the force gallery (Fig.~\ref{fig:gallery}).  Two important features to note in the incremental work fluctuations are: sample-to-sample variation is quite high (e.g.~small fluctuations for trajectory (d) and large fluctuations for (e)); and fluctuations in time also vary even within a sample (e.g.~trajectory (f) exhibits small fluctuations initially that increase in amplitude as time advances).  This suggests that averaging over (1) samples and/or (2) time will hide variations in dissipation.

The corresponding accumulated work along each trajectory is shown in Fig.~\ref{fig:work}(right) from numerical integration of the work increments.  Here we observe that the work done by the optical trap, $W_{\mathrm{OT}}$ fluctuates positive/negative as expected and also varies significantly between trajectories. We note the net work done by the optical trap on average is zero, since the trapping potential does not vary in time.  Perhaps more interesting because it is directly comparable to fluid mechanics studies~\cite{guasto2010oscillatory,leptos2009dynamics} is the work done by friction, $W_{\mathrm{fr}}$, which is a fluctuating negative quantity that accumulates over time.  At the trajectory level it is clear $W_{\mathrm{fr}}$ also varies between swimmers and over time.  Quantifying the activity of the individual swimmers from the average slope of the $W_{\mathrm{fr}}$ curves in Fig.~\ref{fig:work}(right) shows a range of power dissipation, where the least active swimmer (d) dissipates $0.5 \times 10^6$ $k_B T/$s ($\sim2$ fW) and the most active (e) dissipates $1.9 \times 10^6$ $k_B T/$s ($\sim8$ fW), during the observation window.  The average over all swimmers (a-f) estimates the average power dissipation from viscous forces to be $\langle \vert P_{\mathrm{act}} \vert \rangle = 1.2 \times 10^6 k_B T/$s ($\sim5$ fW).  This average value is very close to the average power dissipated by a \emph{Chlamydomonas} from viscous dissipation, $\sim4$ fW~\cite{guasto2010oscillatory}.  If we calculate the instantaneous power dissipated by friction, we see peak values of $\sim10-15$ fW, which is again in striking agreement with the value of $\sim15$ fW measured by viscous dissipation~\cite{guasto2010oscillatory}.

Overall, this agreement between the optical trap and fluid mechanics approach is quite striking.  In both cases we assume that the mechanical energy generated by the swimmer is dissipated by friction into the viscous fluid; however, our calculation methods are very different.  The fluids studies calculate the power transferred from a \emph{Chlamydomonas} to the viscous fluid from the velocity field gradient, $P = \int 2\mu (\mathbf{\Gamma}:\mathbf{\Gamma})h\,dA$, where $\mu$ is the fluid viscosity, $h$ the fluid height, and $\mathbf{\Gamma}=\frac{1}{2}[\nabla \mathbf{v} + (\nabla \mathbf{v})^T]$ is the rate of strain tensor, $\mathbf{v}$ is the velocity field, and $dA$ is the differential area element. This method inherently involves regularization (smoothing) of the velocity field in space and time from the particle image velocimetry ~\cite{guasto2010oscillatory, ouellette2006quantitative}.  In our study, we calculate the instantaneous power dissipated by friction from the basic definition of power, $P = \mathbf{F}_{\mathrm{fr}} \circ \dot{\mathbf{r}}$, where $\mathbf{F}_{\mathrm{fr}} = \gamma \dot{\mathbf{r}}$, and $\dot{\mathbf{r}}$ is the instantaneous velocity of the swimmer calculated from the displacement ($\Delta \mathbf{r}$) and time ($\Delta t$) via back focal plane interferometry. In both approaches, low-Reynolds Stokes friction is assumed. The force measurement approach provides several benefits: increased resolution not limited by image acquisition,  model-independent force fluctuations, access to thermal fluctuations, and comparison to stochastic particle models widely used in active matter and stochastic thermodynamics.  From this analysis of force trajectories we observe two things when considering individual trajectories in the time domain:~(1)~for a \emph{Chlamydomonas}, measurement of the fluid velocity field surrounding the swimmer and direct measurement of the net force fluctuations of an active swimmer lead to similar estimates of the power-dissipation due to swimming; (2)~there is significant variation in the work fluctuations in time as well as between different swimmers.

\begin{figure*}
\includegraphics[width=1\textwidth]{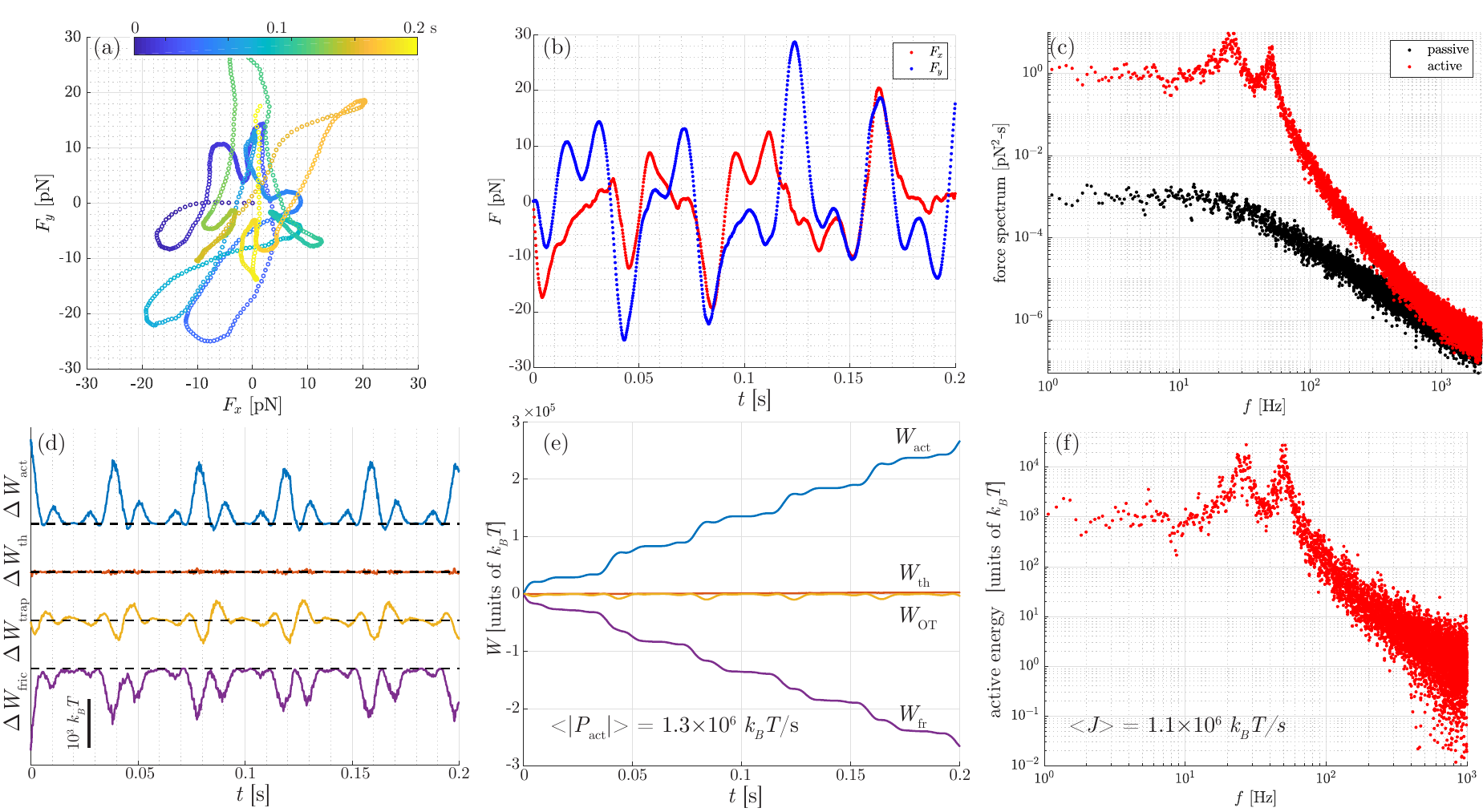}
\caption{\label{fig:sims} Simulations of the Active Beating Swimmer (ABS) model.  Adding force oscillations to the equation of motion recapitulates the main features of our experimental measurements.  This includes:  (a) complex force trajectories oscillating around the trap center;  (b) force oscillations in time; (c) local peaks in the force spectrum; (d) fluctuating work increments; (e) comparable dissipation by friction; and (f) local peaks in the dissipation spectrum. }
\end{figure*}

\subsection{Numerical simulations of an oscillating swimmer}

One glaring deficiency of the analytic model is the inability to capture the local peaks in the force spectrum (Fig.~\ref{fig:PSD}).  This is not surprising since the paradigmatic self-propelled particle models (AOUP and ABP) exhibit exponentially decaying active force correlations~\cite{fodor2018statistical}, with no oscillations.  However, a microswimmer, such as a \emph{Chlamydomonas},  cyclically beats its flagella to swim and thus the active forces generated must be more complex.  In an effort to explore this behavior we modify the ABP model~\cite{romanczuk2012active} to include an oscillating force and use numerical simulations to analyze the resulting force dynamics. We refer to this model for an active force as the Active Beating Swimmer (ABS) model of a self-propelled particle.  Similar models have been developed and explored in the absence of a trapping potential~\cite{babel2014swimming}.

To investigate the ABS model we compute trajectories using equation \ref{eqn:DiscreteTrajectory} for the discrete self-propulsion velocity
\begin{equation}
    \mathbf{u}_i = \mathbf{v}_{0}\left(1 + \sum\displaylimits_{k=1}^N \lambda_k\cos(2\pi f_k t_i)\right),
\end{equation}
where $\mathbf{v}_{0} = (v_0\cos\theta_i,v_0\sin\theta_i)$ is a characteristic velocity with mean speed $v_0$ and a stochastic angle of orientation $\theta_i$ diffusing from $\theta_0 = 0$ with a rotational diffusion coefficient $D_R$; $\lambda_k$ determines the amplitude of the velocity of term $k$; $f_k$ is the frequency; and $N$ indicates the number of oscillatory terms applied. In our case, we choose to represent the ABS model as
\begin{equation}
    \mathbf{u}_i =\mathbf{v}_{0}\left(1 + \lambda\cos(2\pi f_1 t_i)+\lambda\cos(2\pi f_2 t_i)\right),
    \label{eqn:PropulsionVelocity}
\end{equation}
and extract $v_{0}$, $\lambda$, and $f_k$ from experiments. We extract $\langle \vert v_0 \vert \rangle = 38$ $\mu$m/s from fitting equation \ref{eqn:psd} (Fig.~\ref{fig:Param}), $f_1 = 25$ Hz and $f_2 = 50$ Hz correspond to frequency peaks in Fig.~\ref{fig:Non-eq}, and both oscillations have the same amplitude of $V = v_0 \lambda = 350$ $\mu$m/s as estimated from the range of forces measured ($V \sim \vert \mathbf{F} \vert / \gamma$) and is consistent with velocities measured previously~\cite{guasto2010oscillatory, kurtuldu2013flagellar,goldstein2015green}.  The rotational diffusion coefficient is $D_R = 1/\tau$, for a persistence time $\tau$ extracted from experimental data (Fig.~\ref{fig:Param}). Equation \ref{eqn:PropulsionVelocity} is placed into equation \ref{eqn:DiscreteTrajectory}, and the numerical trajectory is simulated over 200 seconds with the input parameters shown in Table \ref{tab:Parameters}.
\begin{table}[h]
\centering
\caption{Parameters for Simulation}
	\begin{tabular}{@{}l|c|l@{}}
		\toprule[1.5pt]
		\multicolumn{1}{c|}{\textbf{Parameter}}	&	\textbf{ Symbol }	&	\multicolumn{1}{c}{\textbf{Value}}	\\
		\toprule[1.0pt]
		time step	            &  	$\Delta t$  & 	$2.0\times 10^{-4}\textrm{ s}$				\\
		friction coefficient	&	$\gamma$	&	$6.6\times 10^{-8} \textrm{ kg/s}$			\\
		trap stiffness	        &	$\kappa$	&	$10^{-6} \textrm{ N/m}$				\\
		thermal diffusion coefficient	&  $D$	&	$6.3\times 10^{-14} \textrm{ m$^2$/s}$		\\
		rotational diffusion coefficient & $D_R$  &  25.6 rad$^2$/s \\ 
		active velocity amplitude       & $V$   & $350\times 10^{-6} \textrm{ m/s}$ \\
		active velocity mean            & $v_0$ & $38\times 10^{-6} \textrm{ m/s}$ \\
		frequency one           & $f_1$         & $25 \textrm{ Hz}$ \\
		frequency two           & $f_2$         & 50 \textrm{ Hz} \\
		\bottomrule[1.0pt]
	\end{tabular} 
	\label{tab:Parameters}
\end{table}

The simple ABS model is able to recapitulate the main features of our experiments, including not only the local peaks in the force spectrum and active energy, but also the trajectory-level dynamics of forces and energetics.  Subsequent calculations of the energy dissipation rate from simulated trajectories also agree with experimental values ($\sim5$ fW).  Interestingly, in addition to agreeing with experimental results on average, the trajectory-level dynamics for the ABS simulations also mimic the measurements.  This is first evident in the force trajectories showing complex oscillatory dynamics (Fig.~\ref{fig:sims}a,b), and further in the fluctuations in work shown in (Fig.~\ref{fig:sims}d,e).  These discrete work fluctuations were computed for $\Delta t = 2.0 \times 10^{-4}$ s for direct comparison to experiments with the same time resolution (shown in Fig.~\ref{fig:work}); however, it is important to note that moving to smaller time steps may not be rigorously valid since some of the limiting quantities diverge as $\Delta t \rightarrow 0^+$. Together, these simulations suggest that simple modifications of self-propelled particle models (e.g. adding oscillatory forces) can be used to quantitatively describe the average and trajectory-level dynamics of microswimmers as observed via experiments.

\section{Conclusion}
We have experimentally measured the stochastic force dynamics generated by a \emph{Chlamydomonas} microswimmer in an optical trap using the photon momentum method.  We find that its swimming motion generates complex oscillatory force dynamics including rotational motion.  Using self-propelled particle models we characterize the non-equilibrium activity of microswimmers on average, and use trajectory-level analysis to quantify the energetics of single microswimmers with time.  Overall, using our optical trap measurements and a particle-based Langevin approach we find that \emph{Chlamydomonas} exhibit an average power dissipation of $\langle \vert P_{\mathrm{act}} \vert \rangle = 1.2 \times 10^6 k_B T/$s ($\sim5$ fW), which is in striking agreement with previous fluid mechanics approaches.  The force measurement approach, an alternative to well-established fluid approaches, is well-suited to characterize the stochasticity and fluctuations of microswimmer dynamics and reveals complex force patterns not previously accessible. This approach provides direct access to model-independent force fluctuations, high resolution sampling, and data compatible with particle-based models for investigations of stochastic thermodynamics.

\section{Acknowledgement}
CJ, MG, and RM acknowledge the Black Family Fellowship.  NB and WWA acknowledge funding from the CSUF RSCA.  Part of this work was completed at the QCBNet hackathon supported by NSF grant MCB-1411898.  This material is based upon work supported by the National Science Foundation under Grant No. DMS-2010018.

\appendix

\section{Deriving the finite difference equation}

To simulate individual stochastic trajectories from the equation of motion in \eqref{eqn:EoM}, the velocity is isolated and then integrated over a time interval $[t_{i-1},t_i]$ of size $\Delta t$. The result is
\begin{equation*}
    \mathbf{r}(t_i) - \mathbf{r}(t_{i-1}) = \int\displaylimits_{t_{i-1}}^{t_i}\left(-\frac{\kappa}{\gamma}\mathbf{r} + \mathbf{u} +\sqrt{2D}\boldsymbol\eta\right)dt.
\end{equation*}
which upon formally applying the mean value yields
\begin{equation}
     \mathbf{r}_i - \mathbf{r}_{i-1} = -\frac{\kappa}{\gamma}\mathbf{r}_{i^*}\Delta t + \mathbf{u}_{i^*}\Delta t + \sqrt{2D}\boldsymbol\eta_{i^*}\Delta t,
    \label{eqn:AppendixDeltaPosition}
\end{equation}
for a time $t_{i^*}\in[t_{i-1},t_i]$. In \eqref{eqn:AppendixDeltaPosition}, the subscript notation indicates the evaluation point of corresponding quantity, e.g., $\mathbf{r}_{i^*} \equiv \mathbf{r}(t_{i^*})$. Since equation \eqref{eqn:AppendixDeltaPosition} does not include a stochastic variable multiplying a stochastic increment, the evaluation point within the time interval is of little consequence~\cite{gardiner1985handbook}. The left endpoint is chosen for convenience and consistency, allowing us to write the particle's discrete trajectory as
\begin{equation}
    \mathbf{r}_i = \mathbf{r}_{i-1} - \frac{\kappa}{\gamma}\mathbf{r}_{i-1}\Delta t + \mathbf{u}_{i-1}\Delta t + \sqrt{2D}\,\Delta\mathbf{B}_{i},
\end{equation}
where $\Delta\mathbf{B}_{i}=\boldsymbol\eta_{i-1}\Delta t$ is an increment of the Wiener process. We note that
\begin{equation*}
\Delta \mathbf{B}_{i}\sim\mathcal{N}(0,\Delta t)\sim\sqrt{\Delta t}\,\mathcal{N}(0,1),
\end{equation*}
where $\mathcal{N}(0,1)$ denotes a normal distribution with a mean of zero and a variance of unity. We therefore have reason to rewrite $\Delta \mathbf{B}_{i}$ as
\begin{equation*}
\Delta \mathbf{B}_{i} = \sqrt{\Delta t}\,\mathbf{Z}_{i-1},
\label{eqn:W=Z-}
\end{equation*}
where $\mathbf{Z}_i$ is a normally distributed random variable that satisfies $\langle Z_{\alpha,i}\rangle=0$ and $\langle Z_{\alpha,i} Z_{\beta,j}\rangle=\delta_{\alpha\beta}\delta_{ij}$. The reason for such manipulation is to lower the computational cost. We are left with
\begin{equation}
    \mathbf{r}_i = \mathbf{r}_{i-1} - \frac{\kappa}{\gamma}\mathbf{r}_{i-1}\Delta t + \mathbf{u}_{i-1}\Delta t + \sqrt{2D\Delta t}\,\mathbf{Z}_{i-1}.
    \label{eqn:AppendixDiscreteTrajectory}
\end{equation}

\section{Calculating the power spectral density}

The components $x(t)$ and $y(t)$ of $\mathbf{r}(t)$ solving \eqref{eqn:EoM} satisfy equivalent stochastic differential equations. So without loss of generality, consider the $x$ process, which is exactly given by 
\begin{equation}
\begin{split}
  x(t)& = x_0 e^{-\mu t} + \int_0^t e^{\mu (z - t)} u(z) dz \\ 
  & \qquad + \sqrt{2 D} \int_0^t e^{\mu (z - t)} \eta(z) dz.
\end{split}
\end{equation}
where $\mu = \kappa/\gamma$ and $\eta$ is a standard Guassian white noise process. \

Assuming the mutual independence of the random variables $x_0$, $u$ and $\eta(t)$, the positional autocorrelation 
\[
\begin{split}
    E[x(t)x(s)] 
  & =  e^{-\mu (t+s)}E[{x_0}^2] \\ & + \int_0^t \int_0^s e^{\mu (z - t + \zeta - s)} E[ u(z)  u(\zeta)] d\zeta dz  \qquad  
 \\ & +  2 D \int_0^t\int_0^s e^{\mu (z - t + \zeta - s)} \delta(z-\zeta)   d\zeta dz,
\end{split}
\]
where we have used that $E[\eta(t)] = E[u(t)] = 0$ and that $E[\eta(z) \eta(\zeta)] = \delta(z-\zeta)$.

 With the assumption $t \geq s$, the integral
\[
 \begin{split}
 \int_0^t &\int_0^s e^{\mu (z - t + \zeta - s)} \delta(z-\zeta)   d\zeta dz \\
 & = \int_0^s \int_0^s e^{\mu (z - t + \zeta - s)} \delta(z-\zeta)   d\zeta dz \\
 & = \int_0^s  e^{\mu (2z - t - s)} dz 
 = \frac{\left(e^{2 \mu  s}-1\right) e^{-\mu (s+t)}}{2 \mu }
 \end{split}
\]
Alternately, if $s > t$, a similar calculation shows the integral is equal to the same expression but with  $t$ and $s$ switched. Hence, the positional autocorrelation is
\[
 \begin{split}
    E[x(t)x(s)] 
  & =  e^{-\mu (t+s)} \Big( E[{x_0}^2] 
  -  \frac{D - De^{2 \mu \min\{s,t\}}}{\mu} \Big) \\ 
  & \quad + \int_0^t \int_0^s e^{\mu (z - t + \zeta - s)} E[ u(z)  u(\zeta)] d\zeta dz,  
 \end{split}
\]
for all $t\geq 0$ and $s\geq 0$.

\section{Active Ornstein--Uhlenbeck particles (AOUPs)}

Active Ornstein--Uhlenbeck particles have an internal stochastic velocity $u$ specified by the stochastic differential equation
\begin{equation}
       \dot{u} + \tau^{-1} u = \sqrt{2 A}\tau^{-1} \chi(t), 
\end{equation}
for a Gaussian white noise process $\chi(t)$, persistence time $\tau$ and active diffusion coefficient $A$. Observe that this equation is in the exact same form as \eqref{eqn:EoM} but with $u \equiv 0$, $\gamma = 1$, $\kappa = \tau^{-1}$ and $D = A\tau^{-2}$. Hence, by the previous work, the autocorrelation of self-propulsive velocity $u(t)$ is
\[
    E[u(t)u(s)] =  e^{-(t+s)/\tau} \Big( E[{u_0}^2] 
  - \frac{A - A e^{2\min\{s,t\}/\tau}}{\tau} \Big),
\]
where $u_0$ is the initial velocity of the process that has no directional bias, i.e., $E[u_0] = 0$.
Setting $E[{u_0}^2] = A/\tau$ reduces the previous formula to
\begin{equation}\label{aoups_autocorr}
 E[u(s)u(t)] = \frac{A}{\tau} e^{- \left| t-s\right|/\tau },
\end{equation}
making the process $u$ stationary in the wide sense.

From expression \eqref{aoups_autocorr}, the double integral
\[
 \begin{split}
  \int_0^t & \int_0^s e^{\mu (z - t + \zeta - s)} E[ u(z)  u(\zeta)] d\zeta dz \\
  & = \frac{A e^{-\mu  (t  + s)} }{\tau}\int_0^t \int_0^s e^{\mu (z + \zeta)}  e^{- \left| z - \zeta\right|/\tau } d\zeta dz.
 \end{split}
\]
For simplicity, assume that $s \geq t \geq 0$. Then $s \geq z$, and the interior integral with respect to $\zeta$ can be split into two integrals over $[0,z]$ and $[z,s]$. Specifically,
\[
 \begin{split}
  \int_0^t & \int_0^s e^{\mu (z + \zeta)}  e^{- \left| z - \zeta\right|/\tau } d\zeta dz \\
  & = \int_0^t e^{(\mu - 1/\tau )z} \left(\int_0^z    e^{ (\mu + 1/\tau )\zeta } d\zeta\right) dz \\
  & \qquad \quad + \int_0^t  e^{(\mu + 1/\tau )z}\left(\int_z^s   e^{ (\mu - 1/\tau)\zeta} d\zeta\right) dz\\
  & = \frac{1 - e^{2 \mu t}}{\mu \tau (\mu^2 - 1/\tau^2)} 
  + \frac{e^{\mu(s+t) - (s-t)/\tau}}{\mu^2 - 1/\tau^2}\\
  & \qquad \quad + \frac{1 - e^{(\mu - 1/\tau)s} - e^{ (\mu - 1/\tau)t}}{\mu^2 - 1/\tau^2}.
 \end{split}
\]
If we remove the constraint that $s\geq t \geq 0$, then
\[
 \begin{split}
  \int_0^t & \int_0^s e^{\mu (z + \zeta)}  e^{- \left| z - \zeta\right|/\tau } d\zeta dz \\
  & = \frac{1 - e^{2 \mu \min\{t,s\}}}{\mu \tau (\mu^2 - 1/\tau^2)} 
  + \frac{e^{\mu(s+t) - |s-t|/\tau}}{\mu^2 - 1/\tau^2} \\
  & \qquad \quad + \frac{1 - e^{(\mu - 1/\tau)s} - e^{ (\mu - 1/\tau)t}}{\mu^2 - 1/\tau^2}
 \end{split}
\]
for all $t \geq 0$ and $s \geq 0$, and 
\[
 \begin{split}
   \int_0^t & \int_0^s e^{\mu (z - t + \zeta - s)} E[ u(z)  u(\zeta)] d\zeta dz \\
  & =  \frac{A}{\tau}\frac{e^{-\mu  (t  + s)} - e^{- \mu|s-t|}}{\mu \tau (\mu^2 - 1/\tau^2)} 
  + \frac{A}{\tau}\frac{  e^{ - |s-t|/\tau}}{\mu^2 - 1/\tau^2} \\
  & \qquad \quad +  \frac{A  }{\tau}\frac{e^{-\mu  (t  + s)} - e^{ -( \mu t + s/\tau)} - e^{-(\mu s + t/\tau)}}{\mu^2 - 1/\tau^2}.
 \end{split}
\]

\subsection{Positional Autocorrelation and Power Spectral Density}

After letting $ E[{x_0}^2] = 0$, the previous formula specifies that the autocorrelation of the position
\[
 \begin{split}
  {}  E[x(t)x(s)]
  & = \left(\frac{D}{\mu} - \frac{ A }{\mu \tau^2 (\mu^2 - 1/\tau^2)} \right) e^{- \mu|s-t|}   
    \\ 
  & \quad + \left(\frac{A(\mu + 1/\tau)}{ \mu  \tau(\mu^2 - 1/\tau^2)} -  \frac{D}{\mu}\right)e^{-\mu (t+s)}
  \\ 
  & \quad 
  +  \frac{A  }{\tau}\frac{ e^{ - |s-t|/\tau} - e^{ -( \mu t + s/\tau)} - e^{-(\mu s + t/\tau)}}{\mu^2 - 1/\tau^2}.
 \end{split}
\]

From the definition of the power spectral density of the position $x(t)$, namely
\[
 S_{xx}(\omega)
 = \lim_{T \to \infty} \frac{1}{T}\int_0^T \int_0^T E[x(t)x(s) ] e^{i \omega (t - s)} ds\, dt,
\] 
we can directly integrate the autocorrelation to show
\[
 \begin{split}
  S_{xx}(\omega)
  & = \left(2D   - \frac{2A\tau^{-2}}{\mu^2 - \tau^{-2}} \right) \frac{1}{\mu ^2+\omega ^2} 
    \\
  & \qquad \quad 
  +   \frac{2A\tau^{-2}}{\mu^2 - \tau^{-2} } \frac{1}{\tau^{-2} + \omega^2 }.
 \end{split}
\]
Since the force of the trap in the $x$-direction is $f(x) = -\kappa x$, then its power spectral density 
$S_{\!\!\:\textit{ff}}(\omega) = \kappa^2 S_{xx}(\omega)$, i.e.,
\[
 \begin{split}
  S_{\!\!\:\textit{ff}}(\omega)
  & = \left(2\kappa^2D   - \frac{2A\kappa^2\tau^{-2}}{\mu^2 - \tau^{-2}} \right) \frac{1}{\mu ^2+\omega ^2} 
    \\
  & \qquad \quad 
  +   \frac{2A\kappa^2\tau^{-2}}{\mu^2 - \tau^{-2} } \frac{1}{\tau^{-2} + \omega^2 }.
 \end{split}
\]
Finally, defining $v_0$ as characteristic strength of the active velocity via the formula $A = \tau {v_0}^2$ produces the equation given in \eqref{eqn:psd} for the power spectral density of the optical trap force.

\bibliography{apssamp}

\providecommand{\noopsort}[1]{}\providecommand{\singleletter}[1]{#1}%
\begin{thebibliography}{73}%
\makeatletter
\providecommand \@ifxundefined [1]{%
 \@ifx{#1\undefined}
}%
\providecommand \@ifnum [1]{%
 \ifnum #1\expandafter \@firstoftwo
 \else \expandafter \@secondoftwo
 \fi
}%
\providecommand \@ifx [1]{%
 \ifx #1\expandafter \@firstoftwo
 \else \expandafter \@secondoftwo
 \fi
}%
\providecommand \natexlab [1]{#1}%
\providecommand \enquote  [1]{``#1''}%
\providecommand \bibnamefont  [1]{#1}%
\providecommand \bibfnamefont [1]{#1}%
\providecommand \citenamefont [1]{#1}%
\providecommand \href@noop [0]{\@secondoftwo}%
\providecommand \href [0]{\begingroup \@sanitize@url \@href}%
\providecommand \@href[1]{\@@startlink{#1}\@@href}%
\providecommand \@@href[1]{\endgroup#1\@@endlink}%
\providecommand \@sanitize@url [0]{\catcode `\\12\catcode `\$12\catcode
  `\&12\catcode `\#12\catcode `\^12\catcode `\_12\catcode `\%12\relax}%
\providecommand \@@startlink[1]{}%
\providecommand \@@endlink[0]{}%
\providecommand \url  [0]{\begingroup\@sanitize@url \@url }%
\providecommand \@url [1]{\endgroup\@href {#1}{\urlprefix }}%
\providecommand \urlprefix  [0]{URL }%
\providecommand \Eprint [0]{\href }%
\providecommand \doibase [0]{http://dx.doi.org/}%
\providecommand \selectlanguage [0]{\@gobble}%
\providecommand \bibinfo  [0]{\@secondoftwo}%
\providecommand \bibfield  [0]{\@secondoftwo}%
\providecommand \translation [1]{[#1]}%
\providecommand \BibitemOpen [0]{}%
\providecommand \bibitemStop [0]{}%
\providecommand \bibitemNoStop [0]{.\EOS\space}%
\providecommand \EOS [0]{\spacefactor3000\relax}%
\providecommand \BibitemShut  [1]{\csname bibitem#1\endcsname}%
\let\auto@bib@innerbib\@empty
\bibitem [{\citenamefont {Lauga}\ and\ \citenamefont
  {Goldstein}(2012)}]{lauga2012microswimmers}%
  \BibitemOpen
  \bibfield  {author} {\bibinfo {author} {\bibfnamefont {E.}~\bibnamefont
  {Lauga}}\ and\ \bibinfo {author} {\bibfnamefont {R.~E.}\ \bibnamefont
  {Goldstein}},\ }\href@noop {} {\bibfield  {journal} {\bibinfo  {journal}
  {Phys. Today}\ }\textbf {\bibinfo {volume} {65}},\ \bibinfo {pages} {30}
  (\bibinfo {year} {2012})}\BibitemShut {NoStop}%
\bibitem [{\citenamefont {Purcell}(1977)}]{purcell1977life}%
  \BibitemOpen
  \bibfield  {author} {\bibinfo {author} {\bibfnamefont {E.~M.}\ \bibnamefont
  {Purcell}},\ }\href@noop {} {\bibfield  {journal} {\bibinfo  {journal}
  {American journal of physics}\ }\textbf {\bibinfo {volume} {45}},\ \bibinfo
  {pages} {3} (\bibinfo {year} {1977})}\BibitemShut {NoStop}%
\bibitem [{\citenamefont {Lauga}\ and\ \citenamefont
  {Powers}(2009)}]{lauga2009hydrodynamics}%
  \BibitemOpen
  \bibfield  {author} {\bibinfo {author} {\bibfnamefont {E.}~\bibnamefont
  {Lauga}}\ and\ \bibinfo {author} {\bibfnamefont {T.~R.}\ \bibnamefont
  {Powers}},\ }\href@noop {} {\bibfield  {journal} {\bibinfo  {journal}
  {Reports on Progress in Physics}\ }\textbf {\bibinfo {volume} {72}},\
  \bibinfo {pages} {096601} (\bibinfo {year} {2009})}\BibitemShut {NoStop}%
\bibitem [{\citenamefont {Cisneros}\ \emph {et~al.}(2010)\citenamefont
  {Cisneros}, \citenamefont {Cortez}, \citenamefont {Dombrowski}, \citenamefont
  {Goldstein},\ and\ \citenamefont {Kessler}}]{cisneros2010fluid}%
  \BibitemOpen
  \bibfield  {author} {\bibinfo {author} {\bibfnamefont {L.~H.}\ \bibnamefont
  {Cisneros}}, \bibinfo {author} {\bibfnamefont {R.}~\bibnamefont {Cortez}},
  \bibinfo {author} {\bibfnamefont {C.}~\bibnamefont {Dombrowski}}, \bibinfo
  {author} {\bibfnamefont {R.~E.}\ \bibnamefont {Goldstein}}, \ and\ \bibinfo
  {author} {\bibfnamefont {J.~O.}\ \bibnamefont {Kessler}},\ }in\ \href@noop {}
  {\emph {\bibinfo {booktitle} {Animal Locomotion}}}\ (\bibinfo  {publisher}
  {Springer},\ \bibinfo {year} {2010})\ pp.\ \bibinfo {pages}
  {99--115}\BibitemShut {NoStop}%
\bibitem [{\citenamefont {Elgeti}\ \emph {et~al.}(2015)\citenamefont {Elgeti},
  \citenamefont {Winkler},\ and\ \citenamefont {Gompper}}]{elgeti2015physics}%
  \BibitemOpen
  \bibfield  {author} {\bibinfo {author} {\bibfnamefont {J.}~\bibnamefont
  {Elgeti}}, \bibinfo {author} {\bibfnamefont {R.~G.}\ \bibnamefont {Winkler}},
  \ and\ \bibinfo {author} {\bibfnamefont {G.}~\bibnamefont {Gompper}},\
  }\href@noop {} {\bibfield  {journal} {\bibinfo  {journal} {Reports on
  progress in physics}\ }\textbf {\bibinfo {volume} {78}},\ \bibinfo {pages}
  {056601} (\bibinfo {year} {2015})}\BibitemShut {NoStop}%
\bibitem [{\citenamefont {Guasto}\ \emph {et~al.}(2012)\citenamefont {Guasto},
  \citenamefont {Rusconi},\ and\ \citenamefont {Stocker}}]{guasto2012fluid}%
  \BibitemOpen
  \bibfield  {author} {\bibinfo {author} {\bibfnamefont {J.~S.}\ \bibnamefont
  {Guasto}}, \bibinfo {author} {\bibfnamefont {R.}~\bibnamefont {Rusconi}}, \
  and\ \bibinfo {author} {\bibfnamefont {R.}~\bibnamefont {Stocker}},\
  }\href@noop {} {\bibfield  {journal} {\bibinfo  {journal} {Annual Review of
  Fluid Mechanics}\ }\textbf {\bibinfo {volume} {44}},\ \bibinfo {pages} {373}
  (\bibinfo {year} {2012})}\BibitemShut {NoStop}%
\bibitem [{\citenamefont {Koch}\ and\ \citenamefont
  {Subramanian}(2011)}]{koch2011collective}%
  \BibitemOpen
  \bibfield  {author} {\bibinfo {author} {\bibfnamefont {D.~L.}\ \bibnamefont
  {Koch}}\ and\ \bibinfo {author} {\bibfnamefont {G.}~\bibnamefont
  {Subramanian}},\ }\href@noop {} {\bibfield  {journal} {\bibinfo  {journal}
  {Annual Review of Fluid Mechanics}\ }\textbf {\bibinfo {volume} {43}},\
  \bibinfo {pages} {637} (\bibinfo {year} {2011})}\BibitemShut {NoStop}%
\bibitem [{\citenamefont {Jeanneret}\ \emph {et~al.}(2016)\citenamefont
  {Jeanneret}, \citenamefont {Contino},\ and\ \citenamefont
  {Polin}}]{jeanneret2016brief}%
  \BibitemOpen
  \bibfield  {author} {\bibinfo {author} {\bibfnamefont {R.}~\bibnamefont
  {Jeanneret}}, \bibinfo {author} {\bibfnamefont {M.}~\bibnamefont {Contino}},
  \ and\ \bibinfo {author} {\bibfnamefont {M.}~\bibnamefont {Polin}},\
  }\href@noop {} {\bibfield  {journal} {\bibinfo  {journal} {The European
  Physical Journal Special Topics}\ }\textbf {\bibinfo {volume} {225}},\
  \bibinfo {pages} {2141} (\bibinfo {year} {2016})}\BibitemShut {NoStop}%
\bibitem [{\citenamefont {Goldstein}(2015)}]{goldstein2015green}%
  \BibitemOpen
  \bibfield  {author} {\bibinfo {author} {\bibfnamefont {R.~E.}\ \bibnamefont
  {Goldstein}},\ }\href@noop {} {\bibfield  {journal} {\bibinfo  {journal}
  {Annual review of fluid mechanics}\ }\textbf {\bibinfo {volume} {47}},\
  \bibinfo {pages} {343} (\bibinfo {year} {2015})}\BibitemShut {NoStop}%
\bibitem [{\citenamefont {Harris}(2001)}]{harris2001chlamydomonas}%
  \BibitemOpen
  \bibfield  {author} {\bibinfo {author} {\bibfnamefont {E.~H.}\ \bibnamefont
  {Harris}},\ }\href@noop {} {\bibfield  {journal} {\bibinfo  {journal} {Annual
  review of plant biology}\ }\textbf {\bibinfo {volume} {52}},\ \bibinfo
  {pages} {363} (\bibinfo {year} {2001})}\BibitemShut {NoStop}%
\bibitem [{\citenamefont {Harris}(2009)}]{harris2009chlamydomonas}%
  \BibitemOpen
  \bibfield  {author} {\bibinfo {author} {\bibfnamefont {E.~H.}\ \bibnamefont
  {Harris}},\ }\href@noop {} {\emph {\bibinfo {title} {The Chlamydomonas
  Sourcebook: Introduction to Chlamydomonas and Its Laboratory Use: Volume
  1}}},\ Vol.~\bibinfo {volume} {1}\ (\bibinfo  {publisher} {Academic press},\
  \bibinfo {year} {2009})\BibitemShut {NoStop}%
\bibitem [{\citenamefont {Merchant}\ \emph {et~al.}(2007)\citenamefont
  {Merchant}, \citenamefont {Prochnik}, \citenamefont {Vallon}, \citenamefont
  {Harris}, \citenamefont {Karpowicz}, \citenamefont {Witman}, \citenamefont
  {Terry}, \citenamefont {Salamov}, \citenamefont {Fritz-Laylin}, \citenamefont
  {Mar{\'e}chal-Drouard} \emph {et~al.}}]{merchant2007chlamydomonas}%
  \BibitemOpen
  \bibfield  {author} {\bibinfo {author} {\bibfnamefont {S.~S.}\ \bibnamefont
  {Merchant}}, \bibinfo {author} {\bibfnamefont {S.~E.}\ \bibnamefont
  {Prochnik}}, \bibinfo {author} {\bibfnamefont {O.}~\bibnamefont {Vallon}},
  \bibinfo {author} {\bibfnamefont {E.~H.}\ \bibnamefont {Harris}}, \bibinfo
  {author} {\bibfnamefont {S.~J.}\ \bibnamefont {Karpowicz}}, \bibinfo {author}
  {\bibfnamefont {G.~B.}\ \bibnamefont {Witman}}, \bibinfo {author}
  {\bibfnamefont {A.}~\bibnamefont {Terry}}, \bibinfo {author} {\bibfnamefont
  {A.}~\bibnamefont {Salamov}}, \bibinfo {author} {\bibfnamefont {L.~K.}\
  \bibnamefont {Fritz-Laylin}}, \bibinfo {author} {\bibfnamefont
  {L.}~\bibnamefont {Mar{\'e}chal-Drouard}},  \emph {et~al.},\ }\href@noop {}
  {\bibfield  {journal} {\bibinfo  {journal} {Science}\ }\textbf {\bibinfo
  {volume} {318}},\ \bibinfo {pages} {245} (\bibinfo {year}
  {2007})}\BibitemShut {NoStop}%
\bibitem [{\citenamefont {Pedley}\ and\ \citenamefont
  {Kessler}(1992)}]{pedley1992hydrodynamic}%
  \BibitemOpen
  \bibfield  {author} {\bibinfo {author} {\bibfnamefont {T.}~\bibnamefont
  {Pedley}}\ and\ \bibinfo {author} {\bibfnamefont {J.~O.}\ \bibnamefont
  {Kessler}},\ }\href@noop {} {\bibfield  {journal} {\bibinfo  {journal}
  {Annual Review of Fluid Mechanics}\ }\textbf {\bibinfo {volume} {24}},\
  \bibinfo {pages} {313} (\bibinfo {year} {1992})}\BibitemShut {NoStop}%
\bibitem [{\citenamefont {Polin}\ \emph {et~al.}(2009)\citenamefont {Polin},
  \citenamefont {Tuval}, \citenamefont {Drescher}, \citenamefont {Gollub},\
  and\ \citenamefont {Goldstein}}]{polin2009chlamydomonas}%
  \BibitemOpen
  \bibfield  {author} {\bibinfo {author} {\bibfnamefont {M.}~\bibnamefont
  {Polin}}, \bibinfo {author} {\bibfnamefont {I.}~\bibnamefont {Tuval}},
  \bibinfo {author} {\bibfnamefont {K.}~\bibnamefont {Drescher}}, \bibinfo
  {author} {\bibfnamefont {J.~P.}\ \bibnamefont {Gollub}}, \ and\ \bibinfo
  {author} {\bibfnamefont {R.~E.}\ \bibnamefont {Goldstein}},\ }\href@noop {}
  {\bibfield  {journal} {\bibinfo  {journal} {Science}\ }\textbf {\bibinfo
  {volume} {325}},\ \bibinfo {pages} {487} (\bibinfo {year}
  {2009})}\BibitemShut {NoStop}%
\bibitem [{\citenamefont {Drescher}\ \emph {et~al.}(2010)\citenamefont
  {Drescher}, \citenamefont {Goldstein}, \citenamefont {Michel}, \citenamefont
  {Polin},\ and\ \citenamefont {Tuval}}]{drescher2010direct}%
  \BibitemOpen
  \bibfield  {author} {\bibinfo {author} {\bibfnamefont {K.}~\bibnamefont
  {Drescher}}, \bibinfo {author} {\bibfnamefont {R.~E.}\ \bibnamefont
  {Goldstein}}, \bibinfo {author} {\bibfnamefont {N.}~\bibnamefont {Michel}},
  \bibinfo {author} {\bibfnamefont {M.}~\bibnamefont {Polin}}, \ and\ \bibinfo
  {author} {\bibfnamefont {I.}~\bibnamefont {Tuval}},\ }\href@noop {}
  {\bibfield  {journal} {\bibinfo  {journal} {Physical Review Letters}\
  }\textbf {\bibinfo {volume} {105}},\ \bibinfo {pages} {168101} (\bibinfo
  {year} {2010})}\BibitemShut {NoStop}%
\bibitem [{\citenamefont {Bayly}\ \emph {et~al.}(2011)\citenamefont {Bayly},
  \citenamefont {Lewis}, \citenamefont {Ranz}, \citenamefont {Okamoto},
  \citenamefont {Pless},\ and\ \citenamefont {Dutcher}}]{bayly2011propulsive}%
  \BibitemOpen
  \bibfield  {author} {\bibinfo {author} {\bibfnamefont {P.}~\bibnamefont
  {Bayly}}, \bibinfo {author} {\bibfnamefont {B.}~\bibnamefont {Lewis}},
  \bibinfo {author} {\bibfnamefont {E.}~\bibnamefont {Ranz}}, \bibinfo {author}
  {\bibfnamefont {R.}~\bibnamefont {Okamoto}}, \bibinfo {author} {\bibfnamefont
  {R.}~\bibnamefont {Pless}}, \ and\ \bibinfo {author} {\bibfnamefont
  {S.}~\bibnamefont {Dutcher}},\ }\href@noop {} {\bibfield  {journal} {\bibinfo
   {journal} {Biophysical journal}\ }\textbf {\bibinfo {volume} {100}},\
  \bibinfo {pages} {2716} (\bibinfo {year} {2011})}\BibitemShut {NoStop}%
\bibitem [{\citenamefont {Kurtuldu}\ \emph {et~al.}(2013)\citenamefont
  {Kurtuldu}, \citenamefont {Tam}, \citenamefont {Hosoi}, \citenamefont
  {Johnson},\ and\ \citenamefont {Gollub}}]{kurtuldu2013flagellar}%
  \BibitemOpen
  \bibfield  {author} {\bibinfo {author} {\bibfnamefont {H.}~\bibnamefont
  {Kurtuldu}}, \bibinfo {author} {\bibfnamefont {D.}~\bibnamefont {Tam}},
  \bibinfo {author} {\bibfnamefont {A.}~\bibnamefont {Hosoi}}, \bibinfo
  {author} {\bibfnamefont {K.~A.}\ \bibnamefont {Johnson}}, \ and\ \bibinfo
  {author} {\bibfnamefont {J.}~\bibnamefont {Gollub}},\ }\href@noop {}
  {\bibfield  {journal} {\bibinfo  {journal} {Physical Review E}\ }\textbf
  {\bibinfo {volume} {88}},\ \bibinfo {pages} {013015} (\bibinfo {year}
  {2013})}\BibitemShut {NoStop}%
\bibitem [{\citenamefont {Leptos}\ \emph {et~al.}(2009)\citenamefont {Leptos},
  \citenamefont {Guasto}, \citenamefont {Gollub}, \citenamefont {Pesci},\ and\
  \citenamefont {Goldstein}}]{leptos2009dynamics}%
  \BibitemOpen
  \bibfield  {author} {\bibinfo {author} {\bibfnamefont {K.~C.}\ \bibnamefont
  {Leptos}}, \bibinfo {author} {\bibfnamefont {J.~S.}\ \bibnamefont {Guasto}},
  \bibinfo {author} {\bibfnamefont {J.~P.}\ \bibnamefont {Gollub}}, \bibinfo
  {author} {\bibfnamefont {A.~I.}\ \bibnamefont {Pesci}}, \ and\ \bibinfo
  {author} {\bibfnamefont {R.~E.}\ \bibnamefont {Goldstein}},\ }\href@noop {}
  {\bibfield  {journal} {\bibinfo  {journal} {Physical Review Letters}\
  }\textbf {\bibinfo {volume} {103}},\ \bibinfo {pages} {198103} (\bibinfo
  {year} {2009})}\BibitemShut {NoStop}%
\bibitem [{\citenamefont {Brun-Cosme-Bruny}\ \emph {et~al.}(2019)\citenamefont
  {Brun-Cosme-Bruny}, \citenamefont {Bertin}, \citenamefont {Coasne},
  \citenamefont {Peyla},\ and\ \citenamefont {Rafa{\"\i}}}]{brun2019effective}%
  \BibitemOpen
  \bibfield  {author} {\bibinfo {author} {\bibfnamefont {M.}~\bibnamefont
  {Brun-Cosme-Bruny}}, \bibinfo {author} {\bibfnamefont {E.}~\bibnamefont
  {Bertin}}, \bibinfo {author} {\bibfnamefont {B.}~\bibnamefont {Coasne}},
  \bibinfo {author} {\bibfnamefont {P.}~\bibnamefont {Peyla}}, \ and\ \bibinfo
  {author} {\bibfnamefont {S.}~\bibnamefont {Rafa{\"\i}}},\ }\href@noop {}
  {\bibfield  {journal} {\bibinfo  {journal} {The Journal of chemical physics}\
  }\textbf {\bibinfo {volume} {150}},\ \bibinfo {pages} {104901} (\bibinfo
  {year} {2019})}\BibitemShut {NoStop}%
\bibitem [{\citenamefont {Brumley}\ \emph {et~al.}(2014)\citenamefont
  {Brumley}, \citenamefont {Wan}, \citenamefont {Polin},\ and\ \citenamefont
  {Goldstein}}]{brumley2014flagellar}%
  \BibitemOpen
  \bibfield  {author} {\bibinfo {author} {\bibfnamefont {D.~R.}\ \bibnamefont
  {Brumley}}, \bibinfo {author} {\bibfnamefont {K.~Y.}\ \bibnamefont {Wan}},
  \bibinfo {author} {\bibfnamefont {M.}~\bibnamefont {Polin}}, \ and\ \bibinfo
  {author} {\bibfnamefont {R.~E.}\ \bibnamefont {Goldstein}},\ }\href@noop {}
  {\bibfield  {journal} {\bibinfo  {journal} {Elife}\ }\textbf {\bibinfo
  {volume} {3}},\ \bibinfo {pages} {e02750} (\bibinfo {year}
  {2014})}\BibitemShut {NoStop}%
\bibitem [{\citenamefont {Niedermayer}\ \emph {et~al.}(2008)\citenamefont
  {Niedermayer}, \citenamefont {Eckhardt},\ and\ \citenamefont
  {Lenz}}]{niedermayer2008synchronization}%
  \BibitemOpen
  \bibfield  {author} {\bibinfo {author} {\bibfnamefont {T.}~\bibnamefont
  {Niedermayer}}, \bibinfo {author} {\bibfnamefont {B.}~\bibnamefont
  {Eckhardt}}, \ and\ \bibinfo {author} {\bibfnamefont {P.}~\bibnamefont
  {Lenz}},\ }\href@noop {} {\bibfield  {journal} {\bibinfo  {journal} {Chaos:
  An Interdisciplinary Journal of Nonlinear Science}\ }\textbf {\bibinfo
  {volume} {18}},\ \bibinfo {pages} {037128} (\bibinfo {year}
  {2008})}\BibitemShut {NoStop}%
\bibitem [{\citenamefont {Uchida}\ and\ \citenamefont
  {Golestanian}(2011)}]{uchida2011generic}%
  \BibitemOpen
  \bibfield  {author} {\bibinfo {author} {\bibfnamefont {N.}~\bibnamefont
  {Uchida}}\ and\ \bibinfo {author} {\bibfnamefont {R.}~\bibnamefont
  {Golestanian}},\ }\href@noop {} {\bibfield  {journal} {\bibinfo  {journal}
  {Physical Review Letters}\ }\textbf {\bibinfo {volume} {106}},\ \bibinfo
  {pages} {058104} (\bibinfo {year} {2011})}\BibitemShut {NoStop}%
\bibitem [{\citenamefont {Wan}\ \emph {et~al.}(2014)\citenamefont {Wan},
  \citenamefont {Leptos},\ and\ \citenamefont {Goldstein}}]{wan2014lag}%
  \BibitemOpen
  \bibfield  {author} {\bibinfo {author} {\bibfnamefont {K.~Y.}\ \bibnamefont
  {Wan}}, \bibinfo {author} {\bibfnamefont {K.~C.}\ \bibnamefont {Leptos}}, \
  and\ \bibinfo {author} {\bibfnamefont {R.~E.}\ \bibnamefont {Goldstein}},\
  }\href@noop {} {\bibfield  {journal} {\bibinfo  {journal} {Journal of the
  Royal Society Interface}\ }\textbf {\bibinfo {volume} {11}},\ \bibinfo
  {pages} {20131160} (\bibinfo {year} {2014})}\BibitemShut {NoStop}%
\bibitem [{\citenamefont {Guo}\ \emph {et~al.}(2020)\citenamefont {Guo},
  \citenamefont {Man}, \citenamefont {Wan},\ and\ \citenamefont
  {Kanso}}]{guo2020intracellular}%
  \BibitemOpen
  \bibfield  {author} {\bibinfo {author} {\bibfnamefont {H.}~\bibnamefont
  {Guo}}, \bibinfo {author} {\bibfnamefont {Y.}~\bibnamefont {Man}}, \bibinfo
  {author} {\bibfnamefont {K.~Y.}\ \bibnamefont {Wan}}, \ and\ \bibinfo
  {author} {\bibfnamefont {E.}~\bibnamefont {Kanso}},\ }\href@noop {}
  {\bibfield  {journal} {\bibinfo  {journal} {arXiv preprint arXiv:2008.07626}\
  } (\bibinfo {year} {2020})}\BibitemShut {NoStop}%
\bibitem [{\citenamefont {Wan}\ and\ \citenamefont
  {Goldstein}(2018)}]{wan2018time}%
  \BibitemOpen
  \bibfield  {author} {\bibinfo {author} {\bibfnamefont {K.~Y.}\ \bibnamefont
  {Wan}}\ and\ \bibinfo {author} {\bibfnamefont {R.~E.}\ \bibnamefont
  {Goldstein}},\ }\href@noop {} {\bibfield  {journal} {\bibinfo  {journal}
  {Physical review letters}\ }\textbf {\bibinfo {volume} {121}},\ \bibinfo
  {pages} {058103} (\bibinfo {year} {2018})}\BibitemShut {NoStop}%
\bibitem [{\citenamefont {Lighthill}(1952)}]{lighthill1952squirming}%
  \BibitemOpen
  \bibfield  {author} {\bibinfo {author} {\bibfnamefont {M.}~\bibnamefont
  {Lighthill}},\ }\href@noop {} {\bibfield  {journal} {\bibinfo  {journal}
  {Communications on Pure and Applied Mathematics}\ }\textbf {\bibinfo {volume}
  {5}},\ \bibinfo {pages} {109} (\bibinfo {year} {1952})}\BibitemShut {NoStop}%
\bibitem [{\citenamefont {Hancock}(1953)}]{hancock1953self}%
  \BibitemOpen
  \bibfield  {author} {\bibinfo {author} {\bibfnamefont {G.}~\bibnamefont
  {Hancock}},\ }\href@noop {} {\bibfield  {journal} {\bibinfo  {journal}
  {Proceedings of the Royal Society of London. Series A. Mathematical and
  Physical Sciences}\ }\textbf {\bibinfo {volume} {217}},\ \bibinfo {pages}
  {96} (\bibinfo {year} {1953})}\BibitemShut {NoStop}%
\bibitem [{\citenamefont {Blake}(1971)}]{blake1971note}%
  \BibitemOpen
  \bibfield  {author} {\bibinfo {author} {\bibfnamefont {J.}~\bibnamefont
  {Blake}},\ }in\ \href@noop {} {\emph {\bibinfo {booktitle} {Proc. Camb. Phil.
  Soc}}},\ Vol.~\bibinfo {volume} {70}\ (\bibinfo {year} {1971})\ pp.\ \bibinfo
  {pages} {303--310}\BibitemShut {NoStop}%
\bibitem [{\citenamefont {Ashkin}\ and\ \citenamefont
  {Dziedzic}(1987)}]{ashkin1987optical}%
  \BibitemOpen
  \bibfield  {author} {\bibinfo {author} {\bibfnamefont {A.}~\bibnamefont
  {Ashkin}}\ and\ \bibinfo {author} {\bibfnamefont {J.~M.}\ \bibnamefont
  {Dziedzic}},\ }\href@noop {} {\bibfield  {journal} {\bibinfo  {journal}
  {Science}\ }\textbf {\bibinfo {volume} {235}},\ \bibinfo {pages} {1517}
  (\bibinfo {year} {1987})}\BibitemShut {NoStop}%
\bibitem [{\citenamefont {Ashkin}\ \emph {et~al.}(1987)\citenamefont {Ashkin},
  \citenamefont {Dziedzic},\ and\ \citenamefont {Yamane}}]{ashkin1987optical2}%
  \BibitemOpen
  \bibfield  {author} {\bibinfo {author} {\bibfnamefont {A.}~\bibnamefont
  {Ashkin}}, \bibinfo {author} {\bibfnamefont {J.~M.}\ \bibnamefont
  {Dziedzic}}, \ and\ \bibinfo {author} {\bibfnamefont {T.}~\bibnamefont
  {Yamane}},\ }\href@noop {} {\bibfield  {journal} {\bibinfo  {journal}
  {Nature}\ }\textbf {\bibinfo {volume} {330}},\ \bibinfo {pages} {769}
  (\bibinfo {year} {1987})}\BibitemShut {NoStop}%
\bibitem [{\citenamefont {Block}\ \emph {et~al.}(1989)\citenamefont {Block},
  \citenamefont {Blair},\ and\ \citenamefont {Berg}}]{block1989compliance}%
  \BibitemOpen
  \bibfield  {author} {\bibinfo {author} {\bibfnamefont {S.~M.}\ \bibnamefont
  {Block}}, \bibinfo {author} {\bibfnamefont {D.~F.}\ \bibnamefont {Blair}}, \
  and\ \bibinfo {author} {\bibfnamefont {H.~C.}\ \bibnamefont {Berg}},\
  }\href@noop {} {\bibfield  {journal} {\bibinfo  {journal} {Nature}\ }\textbf
  {\bibinfo {volume} {338}},\ \bibinfo {pages} {514} (\bibinfo {year}
  {1989})}\BibitemShut {NoStop}%
\bibitem [{\citenamefont {Min}\ \emph {et~al.}(2009)\citenamefont {Min},
  \citenamefont {Mears}, \citenamefont {Chubiz}, \citenamefont {Rao},
  \citenamefont {Golding},\ and\ \citenamefont {Chemla}}]{min2009high}%
  \BibitemOpen
  \bibfield  {author} {\bibinfo {author} {\bibfnamefont {T.~L.}\ \bibnamefont
  {Min}}, \bibinfo {author} {\bibfnamefont {P.~J.}\ \bibnamefont {Mears}},
  \bibinfo {author} {\bibfnamefont {L.~M.}\ \bibnamefont {Chubiz}}, \bibinfo
  {author} {\bibfnamefont {C.~V.}\ \bibnamefont {Rao}}, \bibinfo {author}
  {\bibfnamefont {I.}~\bibnamefont {Golding}}, \ and\ \bibinfo {author}
  {\bibfnamefont {Y.~R.}\ \bibnamefont {Chemla}},\ }\href@noop {} {\bibfield
  {journal} {\bibinfo  {journal} {Nature methods}\ }\textbf {\bibinfo {volume}
  {6}},\ \bibinfo {pages} {831} (\bibinfo {year} {2009})}\BibitemShut {NoStop}%
\bibitem [{\citenamefont {Fischer}\ and\ \citenamefont
  {Berg-S{\o}rensen}(2007)}]{fischer2007calibration}%
  \BibitemOpen
  \bibfield  {author} {\bibinfo {author} {\bibfnamefont {M.}~\bibnamefont
  {Fischer}}\ and\ \bibinfo {author} {\bibfnamefont {K.}~\bibnamefont
  {Berg-S{\o}rensen}},\ }\href@noop {} {\bibfield  {journal} {\bibinfo
  {journal} {Journal of Optics A: Pure and Applied Optics}\ }\textbf {\bibinfo
  {volume} {9}},\ \bibinfo {pages} {S239} (\bibinfo {year} {2007})}\BibitemShut
  {NoStop}%
\bibitem [{\citenamefont {Hendricks}\ \emph {et~al.}(2012)\citenamefont
  {Hendricks}, \citenamefont {Holzbaur},\ and\ \citenamefont
  {Goldman}}]{hendricks2012force}%
  \BibitemOpen
  \bibfield  {author} {\bibinfo {author} {\bibfnamefont {A.~G.}\ \bibnamefont
  {Hendricks}}, \bibinfo {author} {\bibfnamefont {E.~L.}\ \bibnamefont
  {Holzbaur}}, \ and\ \bibinfo {author} {\bibfnamefont {Y.~E.}\ \bibnamefont
  {Goldman}},\ }\href@noop {} {\bibfield  {journal} {\bibinfo  {journal}
  {Proceedings of the National Academy of Sciences}\ }\textbf {\bibinfo
  {volume} {109}},\ \bibinfo {pages} {18447} (\bibinfo {year}
  {2012})}\BibitemShut {NoStop}%
\bibitem [{\citenamefont {Blehm}\ \emph {et~al.}(2013)\citenamefont {Blehm},
  \citenamefont {Schroer}, \citenamefont {Trybus}, \citenamefont {Chemla},\
  and\ \citenamefont {Selvin}}]{blehm2013vivo}%
  \BibitemOpen
  \bibfield  {author} {\bibinfo {author} {\bibfnamefont {B.~H.}\ \bibnamefont
  {Blehm}}, \bibinfo {author} {\bibfnamefont {T.~A.}\ \bibnamefont {Schroer}},
  \bibinfo {author} {\bibfnamefont {K.~M.}\ \bibnamefont {Trybus}}, \bibinfo
  {author} {\bibfnamefont {Y.~R.}\ \bibnamefont {Chemla}}, \ and\ \bibinfo
  {author} {\bibfnamefont {P.~R.}\ \bibnamefont {Selvin}},\ }\href@noop {}
  {\bibfield  {journal} {\bibinfo  {journal} {Proceedings of the National
  Academy of Sciences}\ }\textbf {\bibinfo {volume} {110}},\ \bibinfo {pages}
  {3381} (\bibinfo {year} {2013})}\BibitemShut {NoStop}%
\bibitem [{\citenamefont {Ahmed}\ \emph {et~al.}(2018)\citenamefont {Ahmed},
  \citenamefont {Fodor}, \citenamefont {Almonacid}, \citenamefont {Bussonnier},
  \citenamefont {Verlhac}, \citenamefont {Gov}, \citenamefont {Visco},
  \citenamefont {van Wijland},\ and\ \citenamefont {Betz}}]{ahmed2018active}%
  \BibitemOpen
  \bibfield  {author} {\bibinfo {author} {\bibfnamefont {W.~W.}\ \bibnamefont
  {Ahmed}}, \bibinfo {author} {\bibfnamefont {E.}~\bibnamefont {Fodor}},
  \bibinfo {author} {\bibfnamefont {M.}~\bibnamefont {Almonacid}}, \bibinfo
  {author} {\bibfnamefont {M.}~\bibnamefont {Bussonnier}}, \bibinfo {author}
  {\bibfnamefont {M.-H.}\ \bibnamefont {Verlhac}}, \bibinfo {author}
  {\bibfnamefont {N.}~\bibnamefont {Gov}}, \bibinfo {author} {\bibfnamefont
  {P.}~\bibnamefont {Visco}}, \bibinfo {author} {\bibfnamefont
  {F.}~\bibnamefont {van Wijland}}, \ and\ \bibinfo {author} {\bibfnamefont
  {T.}~\bibnamefont {Betz}},\ }\href@noop {} {\bibfield  {journal} {\bibinfo
  {journal} {Biophysical journal}\ }\textbf {\bibinfo {volume} {114}},\
  \bibinfo {pages} {1667} (\bibinfo {year} {2018})}\BibitemShut {NoStop}%
\bibitem [{\citenamefont {Farr{\'e}}\ and\ \citenamefont
  {Montes-Usategui}(2010)}]{farre2010force}%
  \BibitemOpen
  \bibfield  {author} {\bibinfo {author} {\bibfnamefont {A.}~\bibnamefont
  {Farr{\'e}}}\ and\ \bibinfo {author} {\bibfnamefont {M.}~\bibnamefont
  {Montes-Usategui}},\ }\href@noop {} {\bibfield  {journal} {\bibinfo
  {journal} {Optics express}\ }\textbf {\bibinfo {volume} {18}},\ \bibinfo
  {pages} {11955} (\bibinfo {year} {2010})}\BibitemShut {NoStop}%
\bibitem [{\citenamefont {Farr{\'e}}\ \emph {et~al.}(2017)\citenamefont
  {Farr{\'e}}, \citenamefont {Mars{\`a}},\ and\ \citenamefont
  {Montes-Usategui}}]{farre2017beyond}%
  \BibitemOpen
  \bibfield  {author} {\bibinfo {author} {\bibfnamefont {A.}~\bibnamefont
  {Farr{\'e}}}, \bibinfo {author} {\bibfnamefont {F.}~\bibnamefont
  {Mars{\`a}}}, \ and\ \bibinfo {author} {\bibfnamefont {M.}~\bibnamefont
  {Montes-Usategui}},\ }in\ \href@noop {} {\emph {\bibinfo {booktitle} {Optical
  Tweezers}}}\ (\bibinfo  {publisher} {Springer},\ \bibinfo {year} {2017})\
  pp.\ \bibinfo {pages} {41--76}\BibitemShut {NoStop}%
\bibitem [{\citenamefont {Oddershede}(2012)}]{oddershede2012force}%
  \BibitemOpen
  \bibfield  {author} {\bibinfo {author} {\bibfnamefont {L.~B.}\ \bibnamefont
  {Oddershede}},\ }\href@noop {} {\bibfield  {journal} {\bibinfo  {journal}
  {Nature chemical biology}\ }\textbf {\bibinfo {volume} {8}},\ \bibinfo
  {pages} {879} (\bibinfo {year} {2012})}\BibitemShut {NoStop}%
\bibitem [{\citenamefont {Chattopadhyay}\ \emph {et~al.}(2006)\citenamefont
  {Chattopadhyay}, \citenamefont {Moldovan}, \citenamefont {Yeung},\ and\
  \citenamefont {Wu}}]{chattopadhyay2006swimming}%
  \BibitemOpen
  \bibfield  {author} {\bibinfo {author} {\bibfnamefont {S.}~\bibnamefont
  {Chattopadhyay}}, \bibinfo {author} {\bibfnamefont {R.}~\bibnamefont
  {Moldovan}}, \bibinfo {author} {\bibfnamefont {C.}~\bibnamefont {Yeung}}, \
  and\ \bibinfo {author} {\bibfnamefont {X.}~\bibnamefont {Wu}},\ }\href@noop
  {} {\bibfield  {journal} {\bibinfo  {journal} {Proceedings of the National
  Academy of Sciences}\ }\textbf {\bibinfo {volume} {103}},\ \bibinfo {pages}
  {13712} (\bibinfo {year} {2006})}\BibitemShut {NoStop}%
\bibitem [{\citenamefont {Stellamanns}\ \emph {et~al.}(2014)\citenamefont
  {Stellamanns}, \citenamefont {Uppaluri}, \citenamefont {Hochstetter},
  \citenamefont {Heddergott}, \citenamefont {Engstler},\ and\ \citenamefont
  {Pfohl}}]{stellamanns2014optical}%
  \BibitemOpen
  \bibfield  {author} {\bibinfo {author} {\bibfnamefont {E.}~\bibnamefont
  {Stellamanns}}, \bibinfo {author} {\bibfnamefont {S.}~\bibnamefont
  {Uppaluri}}, \bibinfo {author} {\bibfnamefont {A.}~\bibnamefont
  {Hochstetter}}, \bibinfo {author} {\bibfnamefont {N.}~\bibnamefont
  {Heddergott}}, \bibinfo {author} {\bibfnamefont {M.}~\bibnamefont
  {Engstler}}, \ and\ \bibinfo {author} {\bibfnamefont {T.}~\bibnamefont
  {Pfohl}},\ }\href@noop {} {\bibfield  {journal} {\bibinfo  {journal}
  {Scientific reports}\ }\textbf {\bibinfo {volume} {4}},\ \bibinfo {pages}
  {6515} (\bibinfo {year} {2014})}\BibitemShut {NoStop}%
\bibitem [{\citenamefont {McCord}\ \emph {et~al.}(2005)\citenamefont {McCord},
  \citenamefont {Yukich},\ and\ \citenamefont {Bernd}}]{mccord2005analysis}%
  \BibitemOpen
  \bibfield  {author} {\bibinfo {author} {\bibfnamefont {R.~P.}\ \bibnamefont
  {McCord}}, \bibinfo {author} {\bibfnamefont {J.~N.}\ \bibnamefont {Yukich}},
  \ and\ \bibinfo {author} {\bibfnamefont {K.~K.}\ \bibnamefont {Bernd}},\
  }\href@noop {} {\bibfield  {journal} {\bibinfo  {journal} {Cell motility and
  the cytoskeleton}\ }\textbf {\bibinfo {volume} {61}},\ \bibinfo {pages} {137}
  (\bibinfo {year} {2005})}\BibitemShut {NoStop}%
\bibitem [{\citenamefont {F{\"a}llman}\ \emph {et~al.}(2004)\citenamefont
  {F{\"a}llman}, \citenamefont {Schedin}, \citenamefont {Jass}, \citenamefont
  {Andersson}, \citenamefont {Uhlin},\ and\ \citenamefont
  {Axner}}]{fallman2004optical}%
  \BibitemOpen
  \bibfield  {author} {\bibinfo {author} {\bibfnamefont {E.}~\bibnamefont
  {F{\"a}llman}}, \bibinfo {author} {\bibfnamefont {S.}~\bibnamefont
  {Schedin}}, \bibinfo {author} {\bibfnamefont {J.}~\bibnamefont {Jass}},
  \bibinfo {author} {\bibfnamefont {M.}~\bibnamefont {Andersson}}, \bibinfo
  {author} {\bibfnamefont {B.~E.}\ \bibnamefont {Uhlin}}, \ and\ \bibinfo
  {author} {\bibfnamefont {O.}~\bibnamefont {Axner}},\ }\href@noop {}
  {\bibfield  {journal} {\bibinfo  {journal} {Biosensors and Bioelectronics}\
  }\textbf {\bibinfo {volume} {19}},\ \bibinfo {pages} {1429} (\bibinfo {year}
  {2004})}\BibitemShut {NoStop}%
\bibitem [{\citenamefont {Seifert}(2012)}]{seifert2012stochastic}%
  \BibitemOpen
  \bibfield  {author} {\bibinfo {author} {\bibfnamefont {U.}~\bibnamefont
  {Seifert}},\ }\href@noop {} {\bibfield  {journal} {\bibinfo  {journal}
  {Reports on progress in physics}\ }\textbf {\bibinfo {volume} {75}},\
  \bibinfo {pages} {126001} (\bibinfo {year} {2012})}\BibitemShut {NoStop}%
\bibitem [{\citenamefont {Ouellette}\ \emph {et~al.}(2006)\citenamefont
  {Ouellette}, \citenamefont {Xu},\ and\ \citenamefont
  {Bodenschatz}}]{ouellette2006quantitative}%
  \BibitemOpen
  \bibfield  {author} {\bibinfo {author} {\bibfnamefont {N.~T.}\ \bibnamefont
  {Ouellette}}, \bibinfo {author} {\bibfnamefont {H.}~\bibnamefont {Xu}}, \
  and\ \bibinfo {author} {\bibfnamefont {E.}~\bibnamefont {Bodenschatz}},\
  }\href@noop {} {\bibfield  {journal} {\bibinfo  {journal} {Experiments in
  Fluids}\ }\textbf {\bibinfo {volume} {40}},\ \bibinfo {pages} {301} (\bibinfo
  {year} {2006})}\BibitemShut {NoStop}%
\bibitem [{\citenamefont {Pesce}\ \emph {et~al.}(2020)\citenamefont {Pesce},
  \citenamefont {Jones}, \citenamefont {Marag{\`o}},\ and\ \citenamefont
  {Volpe}}]{pesce2020optical}%
  \BibitemOpen
  \bibfield  {author} {\bibinfo {author} {\bibfnamefont {G.}~\bibnamefont
  {Pesce}}, \bibinfo {author} {\bibfnamefont {P.~H.}\ \bibnamefont {Jones}},
  \bibinfo {author} {\bibfnamefont {O.~M.}\ \bibnamefont {Marag{\`o}}}, \ and\
  \bibinfo {author} {\bibfnamefont {G.}~\bibnamefont {Volpe}},\ }\href@noop {}
  {\bibfield  {journal} {\bibinfo  {journal} {The European Physical Journal
  Plus}\ }\textbf {\bibinfo {volume} {135}},\ \bibinfo {pages} {1} (\bibinfo
  {year} {2020})}\BibitemShut {NoStop}%
\bibitem [{\citenamefont {Harada}\ and\ \citenamefont
  {Sasa}(2005)}]{harada2005equality}%
  \BibitemOpen
  \bibfield  {author} {\bibinfo {author} {\bibfnamefont {T.}~\bibnamefont
  {Harada}}\ and\ \bibinfo {author} {\bibfnamefont {S.-i.}\ \bibnamefont
  {Sasa}},\ }\href@noop {} {\bibfield  {journal} {\bibinfo  {journal} {Physical
  review letters}\ }\textbf {\bibinfo {volume} {95}},\ \bibinfo {pages}
  {130602} (\bibinfo {year} {2005})}\BibitemShut {NoStop}%
\bibitem [{\citenamefont {Shinkai}\ and\ \citenamefont
  {Togashi}(2014)}]{shinkai2014energetics}%
  \BibitemOpen
  \bibfield  {author} {\bibinfo {author} {\bibfnamefont {S.}~\bibnamefont
  {Shinkai}}\ and\ \bibinfo {author} {\bibfnamefont {Y.}~\bibnamefont
  {Togashi}},\ }\href@noop {} {\bibfield  {journal} {\bibinfo  {journal} {EPL
  (Europhysics Letters)}\ }\textbf {\bibinfo {volume} {105}},\ \bibinfo {pages}
  {30002} (\bibinfo {year} {2014})}\BibitemShut {NoStop}%
\bibitem [{\citenamefont {Eldeen}\ \emph {et~al.}(2020)\citenamefont {Eldeen},
  \citenamefont {Muoio}, \citenamefont {Blaisdell-Pijuan}, \citenamefont {La},
  \citenamefont {Gomez}, \citenamefont {Vidal},\ and\ \citenamefont
  {Ahmed}}]{eldeen2020quantifying}%
  \BibitemOpen
  \bibfield  {author} {\bibinfo {author} {\bibfnamefont {S.}~\bibnamefont
  {Eldeen}}, \bibinfo {author} {\bibfnamefont {R.}~\bibnamefont {Muoio}},
  \bibinfo {author} {\bibfnamefont {P.}~\bibnamefont {Blaisdell-Pijuan}},
  \bibinfo {author} {\bibfnamefont {N.}~\bibnamefont {La}}, \bibinfo {author}
  {\bibfnamefont {M.}~\bibnamefont {Gomez}}, \bibinfo {author} {\bibfnamefont
  {A.}~\bibnamefont {Vidal}}, \ and\ \bibinfo {author} {\bibfnamefont
  {W.}~\bibnamefont {Ahmed}},\ }\href@noop {} {\bibfield  {journal} {\bibinfo
  {journal} {Soft Matter}\ } (\bibinfo {year} {2020})}\BibitemShut {NoStop}%
\bibitem [{\citenamefont {Fodor}\ and\ \citenamefont
  {Marchetti}(2018)}]{fodor2018statistical}%
  \BibitemOpen
  \bibfield  {author} {\bibinfo {author} {\bibfnamefont {{\'E}.}~\bibnamefont
  {Fodor}}\ and\ \bibinfo {author} {\bibfnamefont {M.~C.}\ \bibnamefont
  {Marchetti}},\ }\href@noop {} {\bibfield  {journal} {\bibinfo  {journal}
  {Physica A: Statistical Mechanics and its Applications}\ }\textbf {\bibinfo
  {volume} {504}},\ \bibinfo {pages} {106} (\bibinfo {year}
  {2018})}\BibitemShut {NoStop}%
\bibitem [{\citenamefont {Sekimoto}(1998)}]{sekimoto1998langevin}%
  \BibitemOpen
  \bibfield  {author} {\bibinfo {author} {\bibfnamefont {K.}~\bibnamefont
  {Sekimoto}},\ }\href@noop {} {\bibfield  {journal} {\bibinfo  {journal}
  {Progress of Theoretical Physics Supplement}\ }\textbf {\bibinfo {volume}
  {130}},\ \bibinfo {pages} {17} (\bibinfo {year} {1998})}\BibitemShut
  {NoStop}%
\bibitem [{\citenamefont {Romanczuk}\ \emph {et~al.}(2012)\citenamefont
  {Romanczuk}, \citenamefont {B{\"a}r}, \citenamefont {Ebeling}, \citenamefont
  {Lindner},\ and\ \citenamefont {Schimansky-Geier}}]{romanczuk2012active}%
  \BibitemOpen
  \bibfield  {author} {\bibinfo {author} {\bibfnamefont {P.}~\bibnamefont
  {Romanczuk}}, \bibinfo {author} {\bibfnamefont {M.}~\bibnamefont {B{\"a}r}},
  \bibinfo {author} {\bibfnamefont {W.}~\bibnamefont {Ebeling}}, \bibinfo
  {author} {\bibfnamefont {B.}~\bibnamefont {Lindner}}, \ and\ \bibinfo
  {author} {\bibfnamefont {L.}~\bibnamefont {Schimansky-Geier}},\ }\href@noop
  {} {\bibfield  {journal} {\bibinfo  {journal} {The European Physical Journal
  Special Topics}\ }\textbf {\bibinfo {volume} {202}},\ \bibinfo {pages} {1}
  (\bibinfo {year} {2012})}\BibitemShut {NoStop}%
\bibitem [{\citenamefont {Volpe}\ \emph {et~al.}(2014)\citenamefont {Volpe},
  \citenamefont {Gigan},\ and\ \citenamefont {Volpe}}]{volpe2014simulation}%
  \BibitemOpen
  \bibfield  {author} {\bibinfo {author} {\bibfnamefont {G.}~\bibnamefont
  {Volpe}}, \bibinfo {author} {\bibfnamefont {S.}~\bibnamefont {Gigan}}, \ and\
  \bibinfo {author} {\bibfnamefont {G.}~\bibnamefont {Volpe}},\ }\href@noop {}
  {\bibfield  {journal} {\bibinfo  {journal} {American Journal of Physics}\
  }\textbf {\bibinfo {volume} {82}},\ \bibinfo {pages} {659} (\bibinfo {year}
  {2014})}\BibitemShut {NoStop}%
\bibitem [{\citenamefont {Higham}(2001)}]{higham2001algorithmic}%
  \BibitemOpen
  \bibfield  {author} {\bibinfo {author} {\bibfnamefont {D.~J.}\ \bibnamefont
  {Higham}},\ }\href@noop {} {\bibfield  {journal} {\bibinfo  {journal} {SIAM
  review}\ }\textbf {\bibinfo {volume} {43}},\ \bibinfo {pages} {525} (\bibinfo
  {year} {2001})}\BibitemShut {NoStop}%
\bibitem [{\citenamefont {Gardiner}\ \emph {et~al.}(1985)\citenamefont
  {Gardiner} \emph {et~al.}}]{gardiner1985handbook}%
  \BibitemOpen
  \bibfield  {author} {\bibinfo {author} {\bibfnamefont {C.~W.}\ \bibnamefont
  {Gardiner}} \emph {et~al.},\ }\href@noop {} {\emph {\bibinfo {title}
  {Handbook of stochastic methods}}},\ Vol.~\bibinfo {volume} {3}\ (\bibinfo
  {publisher} {springer Berlin},\ \bibinfo {year} {1985})\BibitemShut {NoStop}%
\bibitem [{\citenamefont {Gieseler}\ \emph {et~al.}(2020)\citenamefont
  {Gieseler}, \citenamefont {Gomez-Solano}, \citenamefont {Magazz{\`u}},
  \citenamefont {Castillo}, \citenamefont {Garc{\'\i}a}, \citenamefont
  {Gironella-Torrent}, \citenamefont {Viader-Godoy}, \citenamefont {Ritort},
  \citenamefont {Pesce}, \citenamefont {Arzola} \emph
  {et~al.}}]{gieseler2020optical}%
  \BibitemOpen
  \bibfield  {author} {\bibinfo {author} {\bibfnamefont {J.}~\bibnamefont
  {Gieseler}}, \bibinfo {author} {\bibfnamefont {J.~R.}\ \bibnamefont
  {Gomez-Solano}}, \bibinfo {author} {\bibfnamefont {A.}~\bibnamefont
  {Magazz{\`u}}}, \bibinfo {author} {\bibfnamefont {I.~P.}\ \bibnamefont
  {Castillo}}, \bibinfo {author} {\bibfnamefont {L.~P.}\ \bibnamefont
  {Garc{\'\i}a}}, \bibinfo {author} {\bibfnamefont {M.}~\bibnamefont
  {Gironella-Torrent}}, \bibinfo {author} {\bibfnamefont {X.}~\bibnamefont
  {Viader-Godoy}}, \bibinfo {author} {\bibfnamefont {F.}~\bibnamefont
  {Ritort}}, \bibinfo {author} {\bibfnamefont {G.}~\bibnamefont {Pesce}},
  \bibinfo {author} {\bibfnamefont {A.~V.}\ \bibnamefont {Arzola}},  \emph
  {et~al.},\ }\href@noop {} {\bibfield  {journal} {\bibinfo  {journal} {arXiv
  preprint arXiv:2004.05246}\ } (\bibinfo {year} {2020})}\BibitemShut {NoStop}%
\bibitem [{\citenamefont {Jun}\ \emph {et~al.}(2014)\citenamefont {Jun},
  \citenamefont {Tripathy}, \citenamefont {Narayanareddy}, \citenamefont
  {Mattson-Hoss},\ and\ \citenamefont {Gross}}]{jun2014calibration}%
  \BibitemOpen
  \bibfield  {author} {\bibinfo {author} {\bibfnamefont {Y.}~\bibnamefont
  {Jun}}, \bibinfo {author} {\bibfnamefont {S.~K.}\ \bibnamefont {Tripathy}},
  \bibinfo {author} {\bibfnamefont {B.~R.}\ \bibnamefont {Narayanareddy}},
  \bibinfo {author} {\bibfnamefont {M.~K.}\ \bibnamefont {Mattson-Hoss}}, \
  and\ \bibinfo {author} {\bibfnamefont {S.~P.}\ \bibnamefont {Gross}},\
  }\href@noop {} {\bibfield  {journal} {\bibinfo  {journal} {Biophysical
  journal}\ }\textbf {\bibinfo {volume} {107}},\ \bibinfo {pages} {1474}
  (\bibinfo {year} {2014})}\BibitemShut {NoStop}%
\bibitem [{\citenamefont {Welch}(1967)}]{welch1967use}%
  \BibitemOpen
  \bibfield  {author} {\bibinfo {author} {\bibfnamefont {P.}~\bibnamefont
  {Welch}},\ }\href@noop {} {\bibfield  {journal} {\bibinfo  {journal} {IEEE
  Transactions on audio and electroacoustics}\ }\textbf {\bibinfo {volume}
  {15}},\ \bibinfo {pages} {70} (\bibinfo {year} {1967})}\BibitemShut {NoStop}%
\bibitem [{\citenamefont {Coleman}\ and\ \citenamefont
  {Li}(1996)}]{coleman1996interior}%
  \BibitemOpen
  \bibfield  {author} {\bibinfo {author} {\bibfnamefont {T.~F.}\ \bibnamefont
  {Coleman}}\ and\ \bibinfo {author} {\bibfnamefont {Y.}~\bibnamefont {Li}},\
  }\href@noop {} {\bibfield  {journal} {\bibinfo  {journal} {SIAM Journal on
  optimization}\ }\textbf {\bibinfo {volume} {6}},\ \bibinfo {pages} {418}
  (\bibinfo {year} {1996})}\BibitemShut {NoStop}%
\bibitem [{\citenamefont {Racey}\ \emph {et~al.}(1981)\citenamefont {Racey},
  \citenamefont {Hallett},\ and\ \citenamefont {Nickel}}]{racey1981quasi}%
  \BibitemOpen
  \bibfield  {author} {\bibinfo {author} {\bibfnamefont {T.}~\bibnamefont
  {Racey}}, \bibinfo {author} {\bibfnamefont {R.}~\bibnamefont {Hallett}}, \
  and\ \bibinfo {author} {\bibfnamefont {B.}~\bibnamefont {Nickel}},\
  }\href@noop {} {\bibfield  {journal} {\bibinfo  {journal} {Biophysical
  journal}\ }\textbf {\bibinfo {volume} {35}},\ \bibinfo {pages} {557}
  (\bibinfo {year} {1981})}\BibitemShut {NoStop}%
\bibitem [{\citenamefont {Guasto}\ \emph {et~al.}(2010)\citenamefont {Guasto},
  \citenamefont {Johnson},\ and\ \citenamefont
  {Gollub}}]{guasto2010oscillatory}%
  \BibitemOpen
  \bibfield  {author} {\bibinfo {author} {\bibfnamefont {J.~S.}\ \bibnamefont
  {Guasto}}, \bibinfo {author} {\bibfnamefont {K.~A.}\ \bibnamefont {Johnson}},
  \ and\ \bibinfo {author} {\bibfnamefont {J.~P.}\ \bibnamefont {Gollub}},\
  }\href@noop {} {\bibfield  {journal} {\bibinfo  {journal} {Physical review
  letters}\ }\textbf {\bibinfo {volume} {105}},\ \bibinfo {pages} {168102}
  (\bibinfo {year} {2010})}\BibitemShut {NoStop}%
\bibitem [{\citenamefont {Cortese}\ and\ \citenamefont
  {Wan}(2020)}]{cortese2020control}%
  \BibitemOpen
  \bibfield  {author} {\bibinfo {author} {\bibfnamefont {D.}~\bibnamefont
  {Cortese}}\ and\ \bibinfo {author} {\bibfnamefont {K.~Y.}\ \bibnamefont
  {Wan}},\ }\href@noop {} {\bibfield  {journal} {\bibinfo  {journal} {bioRxiv}\
  } (\bibinfo {year} {2020})}\BibitemShut {NoStop}%
\bibitem [{\citenamefont {Wan}\ and\ \citenamefont
  {Goldstein}(2014)}]{wan2014rhythmicity}%
  \BibitemOpen
  \bibfield  {author} {\bibinfo {author} {\bibfnamefont {K.~Y.}\ \bibnamefont
  {Wan}}\ and\ \bibinfo {author} {\bibfnamefont {R.~E.}\ \bibnamefont
  {Goldstein}},\ }\href@noop {} {\bibfield  {journal} {\bibinfo  {journal}
  {Physical Review Letters}\ }\textbf {\bibinfo {volume} {113}},\ \bibinfo
  {pages} {238103} (\bibinfo {year} {2014})}\BibitemShut {NoStop}%
\bibitem [{\citenamefont {Goldstein}\ \emph {et~al.}(2011)\citenamefont
  {Goldstein}, \citenamefont {Polin},\ and\ \citenamefont
  {Tuval}}]{goldstein2011emergence}%
  \BibitemOpen
  \bibfield  {author} {\bibinfo {author} {\bibfnamefont {R.~E.}\ \bibnamefont
  {Goldstein}}, \bibinfo {author} {\bibfnamefont {M.}~\bibnamefont {Polin}}, \
  and\ \bibinfo {author} {\bibfnamefont {I.}~\bibnamefont {Tuval}},\
  }\href@noop {} {\bibfield  {journal} {\bibinfo  {journal} {Physical Review
  Letters}\ }\textbf {\bibinfo {volume} {107}},\ \bibinfo {pages} {148103}
  (\bibinfo {year} {2011})}\BibitemShut {NoStop}%
\bibitem [{\citenamefont {Quaranta}\ \emph {et~al.}(2015)\citenamefont
  {Quaranta}, \citenamefont {Aubin-Tam},\ and\ \citenamefont
  {Tam}}]{quaranta2015hydrodynamics}%
  \BibitemOpen
  \bibfield  {author} {\bibinfo {author} {\bibfnamefont {G.}~\bibnamefont
  {Quaranta}}, \bibinfo {author} {\bibfnamefont {M.-E.}\ \bibnamefont
  {Aubin-Tam}}, \ and\ \bibinfo {author} {\bibfnamefont {D.}~\bibnamefont
  {Tam}},\ }\href@noop {} {\bibfield  {journal} {\bibinfo  {journal} {Physical
  review letters}\ }\textbf {\bibinfo {volume} {115}},\ \bibinfo {pages}
  {238101} (\bibinfo {year} {2015})}\BibitemShut {NoStop}%
\bibitem [{\citenamefont {R{\"u}ffer}\ and\ \citenamefont
  {Nultsch}(1985)}]{ruffer1985high}%
  \BibitemOpen
  \bibfield  {author} {\bibinfo {author} {\bibfnamefont {U.}~\bibnamefont
  {R{\"u}ffer}}\ and\ \bibinfo {author} {\bibfnamefont {W.}~\bibnamefont
  {Nultsch}},\ }\href@noop {} {\bibfield  {journal} {\bibinfo  {journal} {Cell
  Motility}\ }\textbf {\bibinfo {volume} {5}},\ \bibinfo {pages} {251}
  (\bibinfo {year} {1985})}\BibitemShut {NoStop}%
\bibitem [{\citenamefont {Bechinger}\ \emph {et~al.}(2016)\citenamefont
  {Bechinger}, \citenamefont {Di~Leonardo}, \citenamefont {L{\"o}wen},
  \citenamefont {Reichhardt}, \citenamefont {Volpe},\ and\ \citenamefont
  {Volpe}}]{bechinger2016active}%
  \BibitemOpen
  \bibfield  {author} {\bibinfo {author} {\bibfnamefont {C.}~\bibnamefont
  {Bechinger}}, \bibinfo {author} {\bibfnamefont {R.}~\bibnamefont
  {Di~Leonardo}}, \bibinfo {author} {\bibfnamefont {H.}~\bibnamefont
  {L{\"o}wen}}, \bibinfo {author} {\bibfnamefont {C.}~\bibnamefont
  {Reichhardt}}, \bibinfo {author} {\bibfnamefont {G.}~\bibnamefont {Volpe}}, \
  and\ \bibinfo {author} {\bibfnamefont {G.}~\bibnamefont {Volpe}},\
  }\href@noop {} {\bibfield  {journal} {\bibinfo  {journal} {Reviews of Modern
  Physics}\ }\textbf {\bibinfo {volume} {88}},\ \bibinfo {pages} {045006}
  (\bibinfo {year} {2016})}\BibitemShut {NoStop}%
\bibitem [{\citenamefont {Gallet}\ \emph {et~al.}(2009)\citenamefont {Gallet},
  \citenamefont {Arcizet}, \citenamefont {Bohec},\ and\ \citenamefont
  {Richert}}]{gallet2009power}%
  \BibitemOpen
  \bibfield  {author} {\bibinfo {author} {\bibfnamefont {F.}~\bibnamefont
  {Gallet}}, \bibinfo {author} {\bibfnamefont {D.}~\bibnamefont {Arcizet}},
  \bibinfo {author} {\bibfnamefont {P.}~\bibnamefont {Bohec}}, \ and\ \bibinfo
  {author} {\bibfnamefont {A.}~\bibnamefont {Richert}},\ }\href@noop {}
  {\bibfield  {journal} {\bibinfo  {journal} {Soft matter}\ }\textbf {\bibinfo
  {volume} {5}},\ \bibinfo {pages} {2947} (\bibinfo {year} {2009})}\BibitemShut
  {NoStop}%
\bibitem [{\citenamefont {Guo}\ \emph {et~al.}(2014)\citenamefont {Guo},
  \citenamefont {Ehrlicher}, \citenamefont {Jensen}, \citenamefont {Renz},
  \citenamefont {Moore}, \citenamefont {Goldman}, \citenamefont
  {Lippincott-Schwartz}, \citenamefont {Mackintosh},\ and\ \citenamefont
  {Weitz}}]{guo2014probing}%
  \BibitemOpen
  \bibfield  {author} {\bibinfo {author} {\bibfnamefont {M.}~\bibnamefont
  {Guo}}, \bibinfo {author} {\bibfnamefont {A.~J.}\ \bibnamefont {Ehrlicher}},
  \bibinfo {author} {\bibfnamefont {M.~H.}\ \bibnamefont {Jensen}}, \bibinfo
  {author} {\bibfnamefont {M.}~\bibnamefont {Renz}}, \bibinfo {author}
  {\bibfnamefont {J.~R.}\ \bibnamefont {Moore}}, \bibinfo {author}
  {\bibfnamefont {R.~D.}\ \bibnamefont {Goldman}}, \bibinfo {author}
  {\bibfnamefont {J.}~\bibnamefont {Lippincott-Schwartz}}, \bibinfo {author}
  {\bibfnamefont {F.~C.}\ \bibnamefont {Mackintosh}}, \ and\ \bibinfo {author}
  {\bibfnamefont {D.~A.}\ \bibnamefont {Weitz}},\ }\href@noop {} {\bibfield
  {journal} {\bibinfo  {journal} {Cell}\ }\textbf {\bibinfo {volume} {158}},\
  \bibinfo {pages} {822} (\bibinfo {year} {2014})}\BibitemShut {NoStop}%
\bibitem [{\citenamefont {Bohec}\ \emph {et~al.}(2019)\citenamefont {Bohec},
  \citenamefont {Tailleur}, \citenamefont {van Wijland}, \citenamefont
  {Richert},\ and\ \citenamefont {Gallet}}]{bohec2019distribution}%
  \BibitemOpen
  \bibfield  {author} {\bibinfo {author} {\bibfnamefont {P.}~\bibnamefont
  {Bohec}}, \bibinfo {author} {\bibfnamefont {J.}~\bibnamefont {Tailleur}},
  \bibinfo {author} {\bibfnamefont {F.}~\bibnamefont {van Wijland}}, \bibinfo
  {author} {\bibfnamefont {A.}~\bibnamefont {Richert}}, \ and\ \bibinfo
  {author} {\bibfnamefont {F.}~\bibnamefont {Gallet}},\ }\href@noop {}
  {\bibfield  {journal} {\bibinfo  {journal} {Soft matter}\ }\textbf {\bibinfo
  {volume} {15}},\ \bibinfo {pages} {6952} (\bibinfo {year}
  {2019})}\BibitemShut {NoStop}%
\bibitem [{\citenamefont {Fernandez-Rodriguez}\ \emph
  {et~al.}(2020)\citenamefont {Fernandez-Rodriguez}, \citenamefont {Grillo},
  \citenamefont {Alvarez}, \citenamefont {Rathlef}, \citenamefont {Buttinoni},
  \citenamefont {Volpe},\ and\ \citenamefont {Isa}}]{fernandez2020feedback}%
  \BibitemOpen
  \bibfield  {author} {\bibinfo {author} {\bibfnamefont {M.~A.}\ \bibnamefont
  {Fernandez-Rodriguez}}, \bibinfo {author} {\bibfnamefont {F.}~\bibnamefont
  {Grillo}}, \bibinfo {author} {\bibfnamefont {L.}~\bibnamefont {Alvarez}},
  \bibinfo {author} {\bibfnamefont {M.}~\bibnamefont {Rathlef}}, \bibinfo
  {author} {\bibfnamefont {I.}~\bibnamefont {Buttinoni}}, \bibinfo {author}
  {\bibfnamefont {G.}~\bibnamefont {Volpe}}, \ and\ \bibinfo {author}
  {\bibfnamefont {L.}~\bibnamefont {Isa}},\ }\href@noop {} {\bibfield
  {journal} {\bibinfo  {journal} {Nature Communications}\ }\textbf {\bibinfo
  {volume} {11}},\ \bibinfo {pages} {1} (\bibinfo {year} {2020})}\BibitemShut
  {NoStop}%
\bibitem [{\citenamefont {Fodor}\ \emph {et~al.}(2016)\citenamefont {Fodor},
  \citenamefont {Ahmed}, \citenamefont {Almonacid}, \citenamefont {Bussonnier},
  \citenamefont {Gov}, \citenamefont {Verlhac}, \citenamefont {Betz},
  \citenamefont {Visco},\ and\ \citenamefont {van
  Wijland}}]{fodor2016nonequilibrium}%
  \BibitemOpen
  \bibfield  {author} {\bibinfo {author} {\bibfnamefont {{\'E}.}~\bibnamefont
  {Fodor}}, \bibinfo {author} {\bibfnamefont {W.~W.}\ \bibnamefont {Ahmed}},
  \bibinfo {author} {\bibfnamefont {M.}~\bibnamefont {Almonacid}}, \bibinfo
  {author} {\bibfnamefont {M.}~\bibnamefont {Bussonnier}}, \bibinfo {author}
  {\bibfnamefont {N.~S.}\ \bibnamefont {Gov}}, \bibinfo {author} {\bibfnamefont
  {M.-H.}\ \bibnamefont {Verlhac}}, \bibinfo {author} {\bibfnamefont
  {T.}~\bibnamefont {Betz}}, \bibinfo {author} {\bibfnamefont {P.}~\bibnamefont
  {Visco}}, \ and\ \bibinfo {author} {\bibfnamefont {F.}~\bibnamefont {van
  Wijland}},\ }\href@noop {} {\bibfield  {journal} {\bibinfo  {journal} {EPL
  (Europhysics Letters)}\ }\textbf {\bibinfo {volume} {116}},\ \bibinfo {pages}
  {30008} (\bibinfo {year} {2016})}\BibitemShut {NoStop}%
\bibitem [{\citenamefont {Babel}\ \emph {et~al.}(2014)\citenamefont {Babel},
  \citenamefont {Ten~Hagen},\ and\ \citenamefont
  {L{\"o}wen}}]{babel2014swimming}%
  \BibitemOpen
  \bibfield  {author} {\bibinfo {author} {\bibfnamefont {S.}~\bibnamefont
  {Babel}}, \bibinfo {author} {\bibfnamefont {B.}~\bibnamefont {Ten~Hagen}}, \
  and\ \bibinfo {author} {\bibfnamefont {H.}~\bibnamefont {L{\"o}wen}},\
  }\href@noop {} {\bibfield  {journal} {\bibinfo  {journal} {Journal of
  Statistical Mechanics: Theory and Experiment}\ }\textbf {\bibinfo {volume}
  {2014}},\ \bibinfo {pages} {P02011} (\bibinfo {year} {2014})}\BibitemShut
  {NoStop}%
\end{thebibliography}%

\end{document}